%Paper: hep-th/9401115
%From: "Connie Jones, University of Rochester" <CONNIE@URHEP.PAS.ROCHESTER.EDU>
%Date: Mon, 24 Jan 1994 10:47:38 -0500 (EST)

%MATH MODE SYMBOLS
\def\d#1/d#2{ {\partial #1\over\partial #2} }

%ABBREVIATONS
\def\pdr{\partial}

\def\de{\delta}

\def\eps{\epsilon}

\def\half{{1\over 2}}
\def\tr{\hbox{tr}}

\def\un#1{\underline{#1}}

%EQUATION ENVIRONMENTS

\newcount\eqnumber
\def\beq{ \global\advance\eqnumber by 1 $$ }
\def\eeq{ \eqno(\the\eqnumber)$$ }
\def\n{\global\advance \eqnumber by 1\eqno(\the\eqnumber)}
\def\puteqno{
\global\advance \eqnumber by 1 (\the\eqnumber)}
\def\beqs{$$\eqalign}
\def\eeqs{$$}

%REFERENCE AND EQUATION  COUNTERS

\def\ifundefined#1{\expandafter\ifx\csname
#1\endcsname\relax}
          %checks to see if the arguments is the name of a
          %macro already
          %deined.
 \newcount\sectnumber \sectnumber=0
\def\sect#1{ \subsectnumber=0 \advance \sectnumber by 1 {\it
\the \sectnumber
#1} }
\newcount\subsectnumber \subsectnumber=0
\def\subsect#1{ \advance \subsectnumber by 1
{\it \the \sectnumber. \the \subsectnumber #1} }

\newcount\refno \refno=0  %counts references in order of
                         %apperance in text
\def\[#1]{
\ifundefined{#1}
      %define new macro whose expansion is refnumber.
\advance\refno by 1
\expandafter\edef\csname #1\endcsname{\the\refno}\fi[\csname
#1\endcsname]}
\def\refis#1{\noindent\csname #1\endcsname. }

\def\label#1{
\ifundefined{#1}
      %define new macro whose expansion is eqnumber.
\expandafter\edef\csname #1\endcsname{\the\eqnumber}
\else\message{label #1 already in use}
\fi{}}
\def\(#1){(\csname #1\endcsname)}
\def\eqn#1{(\csname #1\endcsname)}

%PREPARATION TO START
\baselineskip=15pt
\parskip=10pt
\magnification=1200
\def\BEGINIGNORE#1ENDIGNORE{}

\baselineskip=20pt

\def\t#1{\tilde{#1}}

          %BEGINNING OF PAPER

\magnification=1200
\def\Lie{\;{\cal L}}
\def\implies{\Rightarrow}
\def\for{\;\;{\rm for}\;\;}
\def\implies{\Rightarrow}
\def\for{\;\;{\rm for}\;\;}
\def\un#1{\underline{#1}}

\def\Lie{\;{\cal L}}
\def\implies{\Rightarrow}
\def\for{\;\;{\rm for}\;\;}
\def\implies{\Rightarrow}
\def\for{\;\;{\rm for}\;\;}
\def\un#1{\underline{#1}}
\def\Hi{\;{\cal H}\;}
\def\sgn{\;{\rm sgn}\;}
\def\htd#1{ \hat{\tilde{#1}} }

          %BEGINNING OF PAPER

\magnification=1200
\def\Lie{\;{\cal L}}
\def\implies{\Rightarrow}
\def\for{\;\;{\rm for}\;\;}
\def\implies{\Rightarrow}
\def\for{\;\;{\rm for}\;\;}
\def\un#1{\underline{#1}}

\centerline{\bf  Quantum Hadrondynamics  in Two  Dimensions}

\centerline{\sl S. G. Rajeev}

 \centerline{Department of Physics and Astronomy,}
\centerline{ University of
Rochester,}
\centerline{Rochester,NY 14627}

\centerline{ABSTRACT}

A nonlocal and nonlinear theory of hadrons, equivalent to the color singlet
sector two dimensional QCD, is constructed. The phase space space of this
theory is an infinite dimensional Grassmannian. The baryon number of QCD
corresponds to a topological invariant  (`virtual rank') of the  Grassmannian.
It is shown that the hadron theory has topological solitons corresponding to
the baryons of QCD. ${1\over N_c}$ plays the role of $\hbar$ in this theory;
$N_c$ must be an integer for topological reasons. We also describe the
quantization of a toy model  with a finite dimensional Grassmannian as the
phase
space. In an appendix, we show that the usual Hartree--Fock
theory of atomic and condensed matter  physics has a natural formulation in
 terms of infinite
dimensional Grassmannians.
\vfill\break

\sect.{Introduction}

Qunatum Chromodynamics (QCD) is the generally accepted
theory
of strong
interactions.  It describes the strongly interacting
particles (hadrons) in terms of their constituents, the quarks and
gluons.
However the quarks and gluons  themselves are not directly
observable; only
their color singlet bound states (hadrons)   exist as
isolated particles.
A challenging problem of particle  physics is to construct a
theory of
strong interactions directly in terms of hadrons. There are
indications
that such a theory exists and is a sort of  string
theory \[strongstring], although a complete understanding
is not available yet.

In this paper we will present such a theory of hadrons (Quantum
Hadrondynamics, QHD) in two spatial space--time dimensions.
It will be
shown that the color singlet sector of two dimensional QCD
(2DQCD) is
equivalent to  QHD for {\it all energy scales and number of
colors}. At low
energies the theory tends to a nonlinear sigma model. For
large $N_c$ it
tends to a  clasical theory; the quantum fluctuations in QHD
are of order
${1\over N_c}$. This classical hadron theory however has no
relation to
classical chromodynamics; it is equivalent instead to the
large $N_c$ limit
of {\it quantum} chromodynamics.

 QHD can be viewed as an extension of the ideas of
Berezin\[berezin],\[perelomov] to QCD. There is also some
previous work by Kikkawa \[kikkawa] in the same spirit,
although our theory differs in some important aspects. One main point
clarified by our work is the origin of baryon number as a
homotopic invariant in QHD.  For this, as well as other reasons, an
understanding of the quadratic constraints satisfied by the field variable is
crucial. The mathematical framework
necessary is the theory of infinite dimensional
Grassmannians
as formulated in the book by Pressley and Segal\[pressegal]. A less explicit
summary of this work was given in \[trieste].

 The main distinguishing feature of QHD is that it is not a
local quantum
field theory. The field variable depends on a pair of space--
time points
separated by a null distance. Also, the theory is highly
nonlinear, the
phase space of the theory being a curved manifold (the
infinite
dimensional Grassmannian). We will first describe the
classical limit of
QHD, which is equivalent to large $N_c$ QCD.
 Small oscillations around  the vacuum describe mesons and
can
be decribed
by a linear approximation to large $N_c$ QCD constructed in
the
work of  't Hooft \[thooft]. Our approach
will construct
the complete theory, which has interactions at every
possible
order, since
the field variable takes values in a curved manifold.
It is important to note that  the infinite $N_c$ limit of
QCD
is equivalent
to an interacting, highly nonlinear, theory. It is only if
we
further
restrict ouselves to  the small oscillations around the
vacuum
that it
becomes a theory of free mesons. If large $N_c$ QCD were a
free
field theory
as is occassionally  claimed there would be no soliton
solutions, which are
necessary to describe baryons as argued by Witten\[witten].
 That baryons should be solitons of a theory
whose small
oscillations are mesons was proposed independently by
Skyrme\[skyrme]
before QCD.  It is now known that this idea is consistent
with
QCD
\[qcdskyrme].
QHD has a homotopically conserved quantum number which can
be
identified
with baryon number. There are  static solutions to our
theory
carrying this
quantum number, which do look like baryons\[rajeevetal].

Our classical hadron theory is  equivalent to the sum over
planar diagrams
of 2DQCD. The field variable $M(x,y)$ of our theory is the
`master field';
we will
obtain its  equations in closed form. It would have been too
hard to obtain
by actually summing diagrams, since there are vortices at
every order. It
is only the geometrical understanding of the phase space as
a
Grassmannian
that makes the formulation of the theory possible.

Quantization of our hadron theory corresponds to studying
QCD
at finite
$N_c$.  In the hadron theory, $N_c$ plays the role of
${1\over
\hbar}$; it
is required to be  a positive integer for homotopic reasons.
The solution
of the  hadron theory in the semiclassical approximation
corresponds to the
${1\over N_c}$ expansion of 2DQCD.

\vfill\eject

\sect.{\it Classical Dynamics on Finite Dimensional
Grassmannians}

\subsect.{\it Parametrizations of the Grassmannian}

We are eventually interested in the infinite dimensional
Grassmanian, which
is the phase space of classical  hadron dynamics. However,
the
particular
infinite dimensional Grassmannian ( the `restricted'
Grassmannian of  G.
Segal \[pressegal]) we need, is very similar to the finite
dimensional
counterparts. It is useful therefore to  study the analogous
finite
dimensional  spaces first. This can be viewed as a
`regularization' of our
theory.

For a positive integer $M$ we will define the Grassmannian
$Gr_M$ to be the
set of all $M\times M$ hermitean matrices satisfying a
quadratic
constraint:
\beq
     Gr_{M}=\{\Phi|\Phi^{\dag}=\Phi;\Phi^2=1\}.
\eeq
The eigenvalues of $\Phi$ will then be $\pm 1$. Each point
$\Phi\in Gr_{M}$
picks out a subspace of $C^{M}$,  the eigenspace of $\Phi$
with eigenvalue
$-1$. Using this correspondence the Grassmannian can be
viewed
as the set of
subspaces of
$C^M$. This is the conventional definition\[chern].

The trace $\tr \Phi$ is an integer:
\beq
     \tr \Phi=M-2m
\eeq
where $m$ is the number of eigenvalues equal to $-1$. Thus
$\tr \Phi$ is
invariant under a continuous deformation of $\Phi$. In fact
$Gr_{M}$  is a
union of  components labelled by $m=1,\cdots M-1$. The case
$m=0$ ($m=M$) is
trivial, containing just one point $\Phi=1(\Phi=-1)$.
\beq
     Gr_{M}=\bigcup_{m=0}^{M} Gr_{m,M}.
\eeq
Each  component can be viewed as a coset space of the
unitary
group:
\beq
     Gr_{m,M}=U(M)/U(m)\times U(M-m).
\eeq
To see this, note that any hermitean matrix can be
diagonalized by a
unitary transformation. There will be precisely $m$
eigenvalues equal to $-1$  and
$M-m$ equal to +1. Thus, for each $\Phi\in Gr_{m,M}$, there
is a $g\in U(M)$ such that \beq
     \Phi=g\eps g^{\dag}
\eeq
where
\beq
     \eps=\pmatrix{-1_{m\times m}&0\cr 0&1_{(M-m)\times (M-
m)}}.
\eeq
Both $g$ and $gh$ correspond to the same $\Phi$ if $h$
commutes with
$\eps$.  The subgroup of elements $U(M)$ that commutes with
$\eps$ is
$U(m)\times U(M-m)$, consisting of unitary matrices  that
are
block
diagonal:
\beq
     h=\pmatrix{h_1&0\cr 0& h_2}.
\eeq
Thus there is a one--one correspondence between $\Phi\in
Gr_{m,M}$ and the
coset space\hfill\break
 $U(M)/U(m)\times U(M-m)$.

 We see now that each component $Gr_{m,M}$  is a connected,
compact
manifold of dimension $2m(M-m)$. Also the map $\Phi\to -
\Phi$
is a
diffeomorphism of $Gr_{m,M}$ to $Gr_{M-m,M}$. If $M=2m$,
this
is a map of
$Gr_{m,M}$ to itself.
The group $U(M)$ acts transitively on each connected
component
$Gr_{m,M}$
by the action
\beq
     \Phi\to g\Phi g^{\dag}.
\eeq

The simplest special case of a Grassmannian is
\beq
     Gr_{1,2}=U(2)/U(1)\times U(1)SU(2)/U(1)=S^2.
\eeq
In fact the description of the general Grassmannian as the
set
of matrices
satisfying a quadratic considtion is similiar to that of
$S^2$
as the set
of vectors of unit length. It is useful to keep this simple
example in mind
as we develop the general theory. More generally, $Gr_{1,M}$
is the complex
projective space $CP^{M-1}$ each point of which is a  one
dimensional of
subspace in $C^M$.

A vector field on the Grassmannian cn be thought of as a
matrix valued
function $V(\Phi)$ satisfying
\beq
     [V(\Phi),\Phi]_+:=V(\Phi)\Phi+\Phi V(\Phi)=0.
\eeq
(If $\Phi=\eps$, this means that $V$ is off--diagonal
$V=\pmatrix{0&v\cr
v^{\dag}&0}$.)  Any such matrix valued function is of the
form
\beq
     V(\Phi)=[\Phi,U(\Phi)]
\eeq
for some other function $U(\Phi)$. However, $U$ is not
uniquely determined
by $V$: the transformation $U(\Phi)\to
U(\Phi)+[\Phi,\Lambda(\Phi)]_+$ will
leave $V(\Phi)$ unchanged. Thus $U(\Phi)$ is a sort of
`potential' for
$V(\Phi)$, since the commutator with $\eps$ is like a
differentiation. (It satisfies the formal rules for cyclic
cohomology; see \[embed]).
If $V(\Phi)$ is a vector field, we have also the
identity
\beq
     [\Phi,[\Phi,V(\Phi)]=4 V(\Phi).
\eeq

\subsect.{Coordinates on the Grassmannian}

It is sometimes convenient to solve the constraint on $\Phi$
by
introducing explicit co--ordinates. Just as for $S^2$, one
needs several
coordinate charts to cover all of the Grassmannian.

Consider first the case of $Gr_{1,2}=CP^1$.  The one
dimensional subspace
picked out by $\Phi$ consists of vectors of type $\lambda
Z^i$, $i=1,2$.
The vector $Z^i$ can be schosen to have unit length, so that
it forms an
`orthonormal  basis' in this one dimensional subspace. Since
$Z^i$ is an
eigenvector of $\Phi$ with eigenvalue $-1$ and the other
eigenvalue is $1$,
\beq
     \Phi_i^j=\delta^j_i-2Z_i^*Z^j.
\eeq
Note that $Z^i\to Z^ih$ will leave $\Phi$ invariant if $h\in
U(1)$.   The
point $\Phi=\eps$ corresponds to $Z^i=\pmatrix{1\cr 0}$.
More
generally, we
can choose the unit vector $Z^i$ in the form
\beq
     Z={1\over \surd(1+ \phi^*\phi)}\pmatrix{1\cr \phi^*}
\eeq
$\phi$ being some complex number.
The magnitude of the first entry is fixed by  the length of
$Z^i$ and its
phase can be chosen as above by an appropriate choice of
$h$.
This gives,
\beq
     \Phi=\pmatrix{1&0\cr 0&1}-{2\over
1+\phi\phi^*}\pmatrix{1&\phi\cr
\phi^*&\phi\phi^*}.
\eeq
One can check easily that $\Phi^2=1$ and $\tr \Phi=0$. This
coordinate
system overs an open neighborhood of $\eps$, but it breaks
down at
$\Phi=\pmatrix{1&0\cr 0&-1}$. Another co--ordinate system,
based on the
choice
\beq
     Z^i={1\over \surd(1+ \phi'^*\phi')}\pmatrix{\phi'\cr 1}
\eeq
and
\beq
     \Phi=\pmatrix{1&0\cr 0&1}-{2\over
1+\phi'\phi'^*}\pmatrix{
                                  \phi'\phi'^*&  \phi^*\cr
\phi'&1\cr}.
\eeq
is well--defined in a neighborhood of this point. The
transformation that
links these systems is
\beq
     \phi'={1\over \phi}.
\eeq
Since the space can be covered by coordinate charts related
by
holomorphic(analytic) cordinate transformations, we see that $Gr_{1,2}$
is a one dimensional  complex manifold.

The above construction can be generalized to an arbitrary
Grassmannian
$Gr_{m,M}$. We will write
\beq
     \Phi_i^j=\delta^j_i-2Z_{ia}^*Z^{ja}.
\eeq
where $Z^{ia}$ for $a=1,\cdots m$ is an orthonormal basis
for the
eigenspace of $\Phi$ with eigenvalues $-1$. $Z$ can be
regarded as an
$M\times m$ matrix and the orthonormality relation is
\beq
     Z^{\dag}Z=1
\eeq
In matrix notation, $\Phi=1-2ZZ^{\dag}$.
$Z$ and $Zh$  give the same $\Phi$ if $h\in U(m)$. We can
choose $Z$ to be
\beq
     Z=\pmatrix{1\cr \phi^{\dag}}[1+\phi\phi^{\dag}]^{-1/2}
\eeq
where $\phi$ is an $ m\times (M-m)$ complex matrix. (The
$1$
in the above
equation stands for an $m\times m$ identity matrix.
$1+\phi\phi^{\dag}$ is
a positive invertible $m\times m$ matrix so it has a well-
defined $-1/2$
power.) This gives
\beq
     \Phi=\pmatrix{1&0\cr 0&1}-
2\pmatrix{(1+\phi\phi^{\dag})^{-
1}&(1+\phi\phi^{\dag})^{-1}\phi\cr
\phi^{\dag}(1+\phi\phi^{\dag})^{-
1}&\phi^{\dag}(1+\phi\phi^{\dag})^{-1}\phi\cr}.
\eeq
One can check explicitly that $\Phi^2=1, \tr \Phi=M-2m$.

The action of $U(M)$ on $\Phi$, $\Phi\to g\Phi g^{\dag}$ can
be seen to be
the transformation
\beq
     \phi\to g\circ \phi= (c+d\phi)(a+b\phi)^{-1}
\eeq
where $g=\pmatrix{a&b\cr c&d\cr}$.

Again this coordinate system breaks down at the points
$\Phi=\pi\eps\pi^{-1}$, $\pi$ being  a
permutation  that interchanges a negative with  a positive
eigenvalue of
$\eps$.
( Permutation of a pir of eigenvectors of $\eps$ is a
unitary
transformation in $C^M$).
If we define,
\beq
     \phi_{\pi}=\pi\circ \phi
\eeq
the  new coordinate  will remain well-defined even at
$\Phi=\pi\eps\pi^{-1}$.
Thus we can cover all of $Gr_{m,M}$ by $\pmatrix{M\cr m}$
coordinate
charts. (This is the number of permutaions that actually
change $\eps$).
Again, since the co--ordinate transformations  are analytic
functions, this
shows that $Gr_{m,M}$ is a complex manifold of dimension
$m(M-
m)$.

\subsect.{The Symplectic Structure on the Grassmannian}

 We will be interested in physical systems for which the
Grassmannian is a
Phase space. Since it is compact,  it cannot be the
cotangent
bundle of any
cofiguration space. It is not possibble to decompose the
dynamical
variables into configuration space and momentum space
variables.  A
canonical formalism is still possible if we have a
symplectic
form;i.e., a
closed, nondegenerate 2-form.  Since each connected
component is a
homogenous space, it is natural to
 look for one that is invariant under the action of $U(M)$.
In
fact there
is a unique (upto overall constant)  homogenous symplectic
form. In terms
our decsription of the Grassmannian in terms of hermitean
matrices,
\beq
     \omega=-{i\over 8}\tr \Phi d\Phi d\Phi.
\eeq
The normalization is chosen so that the integral of $\omega$ over a two--sphere
embedded in the Grassmannian is an integer multiple of $2\pi$. This will be
convenient later.

The form $\omega$  is obviously invariant under the transformation $\Phi\to
g\Phi g^{\dag}$
where $g$ is a constant unitary matrix.
To see that $\omega$  is closed, note that
\beq
     d\omega=-{i\over 8}\tr (d\Phi)^3=-{i\over 8}\tr
(d\Phi)^3\Phi^2
\eeq
since $\Phi^2=1$. By differentiating this constraint,
\beq
     \Phi d\Phi+d\Phi \Phi=0.
\eeq
Now,
\beq
 d\omega=-{i\over 8}\tr\Phi (d\Phi)^3 \Phi={i\over 8}\tr
(d\Phi)^3\Phi^2=-
d\omega
\eeq
so that it is zero.
Since $\omega$  is homogenous, it is enough to verify
nondegeneracy at one
point in each connected component, say $\Phi=\eps\in
Gr_{m,M}$. A tangent
vector $U$ at this point is a hermitean matrix satisfying
\beq
     [\eps,U]_+=0;
\eeq
i.e., of the form
\beq
     U=\pmatrix{0&u\cr u^{\dag}&0}.
\eeq
Then,
\beq
     \omega(U,V)=-{i\over 8}\tr\eps[U,V]=-{i\over
4}\tr[u^{\dag}v-
v^{\dag}u].
\eeq
If $\omega(U,V)=0$  for all $V$ implies that  $u=0$;
$\omega$
is
nondegenerate. Thus $\omega$ is a sympletic form on
$Gr_{m,M}$.

 In fact what we have is a special case of the coadjoint
orbit
construction
of homogenous symplectic manifolds due to Kirillov. The
space
of hemrmitean
matrices is the Lie algebra
of the unitary group. Since the trace is an invariant inner
product, we can
identify  vector space with its dual. The adjoint (identified
with coadjoint)
action   is $\Phi\to g\Phi g^{\dag}$. The Grassmannian
$Gr_{m,M}$ is  just
the orbit of $\eps$.

This suggests a generalization of our construction to more
general coset
spaces, `flag manifolds'. Let $\xi$ be some hermitean matrix
with
eigenvalues $x_\alpha$  with degeneracy $m_{\alpha}$ for
$\alpha=1,\cdots
d$. Clearly $\sum_{\alpha=1}^d m_{\alpha}=M$. The orbit of
$\xi$ is the set
of all elements related to it by an action of $U(M)$:
\beq
     O_{\xi}=\{\Phi=g\xi g^{\dag}|G\in U(M)\}.
\eeq
This is a connected component in the flag manifold
\beq
     Fl(M)=\{\Phi|\Phi^{\dag}=\Phi;\prod_{\alpha=1}^d (\Phi-
x_{\alpha})=0\}.
\eeq
The particular connected component containing $\xi$ is
picked
out by
putting enough constraints on the  traces to determine the
multiplicities:
\beq
     O_{\xi}=\{\Phi|\Phi^{\dag}=\Phi;\prod_{\alpha=1}^d
(\Phi-
x_{\alpha});
 \tr\Phi=\sum m_{\alpha}x_{\alpha};\cdots ;\tr\Phi^{d-1}=
\sum
m_{\alpha}
x_{\alpha}^{d-1}\}
\eeq
It is enough to have $d-1$ trace constraints since $\sum
m_{\alpha}=M$.
As a homogenous space, $O_{\xi}=U(M)/U(m_1)\times\cdots
\times
U(m_{d})$.
The homogenous symplectic form defined by Kirillov's
procedure
is just
\beq
     \omega=-{i\over 8}\tr \Phi (d\Phi)^2.
\eeq
It is clearly invariant under the action of $U(M)$.
At the point $\xi$,
\beq
     \omega(U,V)=-{i\over 8}\tr \xi [U,V]
\eeq
which agrees  with Kirillov's definition \[kirillov] upto a
constant.
It should be interesting to consider the generalization of
the
hadron
theory whose phase space is an infinite dimensional  flag
manifold;
however, it would no longer be equivalent to 2DQCD for
fermionic matter
fields. Perhaps there is some more general quantum field
theory that is
related to it.

\subsect. {\it Poisson Brackets}

We recall some basic facts of classical mechanics to define
the
terminology. \[arnold].
A symplectic form $\omega$ has an inverse $\omega^{-1}$
which
is an
antisymmetric contravariant tensor. Acting on a pair of one-
-
forms, it will
produce a function. In terms of coordinates,
$\omega=\omega_{\mu\nu}dx^{\mu}dx^{\nu}$ and $\omega^{-
1}=\omega^{\mu\nu}\d/d{x^{\mu}} \d/d{x^{\nu}}$,
$\omega^{\mu\nu}$ being
the inverse of the matrix $\omega_{\mu\nu}$.

Observables of a classical dynamical system  are  smooth
functions on a
symplectic manifold, its phase space.
Given a pair of functions $f_1,f_2$, their Poisson Bracket
is
the function
defined by
\beq
     \{f_1,f_2\}=-\omega^{-1}(df_1,df_2).
\eeq
This bracket is  antisymmetric and satisfies the Jacobi
identity if
$\omega$ is closed. The Poisson algebra of a complete set of
observables
defines the symplectic structure uniquely.

To each smooth function $f$ their corresponds a vector
field,
defined by
\beq
     \omega(V_f,.)=df.
\eeq
$V_f$  can be viewed as  the infinitesimal  canonical
transformation
generated by $f$.  Then,
\beq
     \{f_1,f_2\}= \Lie_{V_{f_2}}f_1=\omega(V_{f_1},V_{f_2}).
\eeq
A  classical dynamical system is defined by a manifold , a
symplectic form
$\omega$ and a function $H$, the hamiltonian. The time
evolution of the
system is given by the integral curves of the vector field
$V_H$. Functions
that have zero Poisson bracckets with $H$ are constant under
time evolution.

In our case the matrix elements of $\Phi$ form a complete
set of
observables. Of course they are not all independent, being
related by the
constraints on $\Phi$. It is convenient to think of the
symplectic
structure in terms of the Poisson algebra of these
functions.
For   a
constant hermitean matrix $u$  define the function
\beq
     f_{u}=-\half\tr \Phi u.
\eeq
We can show that
\beq
     \{f_u,f_v\}=f_{-i[u,v]} .
\eeq
Let us first find the symplectic vector field $V_u$
associated
to $f_u$.
Let us think  of the vector field  as a matrix valued
function
$V_u(\Phi)$
satisfying $\Phi V_u(\Phi)+V_u(\Phi)\Phi=0$. Then,
\beq
     \omega(V_u,.)=df_u\implies \quad \tr \left(-{i\over
8}\right)\Phi[V_u(\Phi),d\Phi]=-\half\tr d\Phi u
\eeq
That is,
\beq
     [\Phi,V_u(\Phi)]= -4iu.
\eeq
Using the identity $[\Phi,[\Phi,V_u(\Phi)]]=4V_u(\Phi)$, we
get
\beq
     V_u=-i[u, \Phi].
\eeq
This is just the infinitesimal action of $U(M)$ on $Gr(M)$.
Then,
\beq
     \{f_u,f_v\}(\Phi)=-\Lie_{V_u}f_v(\Phi)=f_{-i[u,v]}(\Phi).
\eeq
Now, let $e^i_j$ be the Weyl matrices,
$(e^i_j)_k^l=\delta^i_k\delta^l_j$.
Then, the matrix elements of $\Phi$ are given by
$\Phi^i_j=\tr
\Phi e^i_j$.
Thus the Poisson algebra of the matrix elements can be
written
as
\beq
     \{\Phi^i_j,\Phi^k_l\}=-2i(\Phi^k_j\delta^i_l-
\Phi^i_l\de^k_j).
\eeq
The Poisson algebra   of a dynamical system whose phase
space is the
Grassmannian can be defined by the above relations on  its
generators along
with the constraints $\Phi^i_j\Phi^j_k=\de^i_k$.

The functions $f_u$ generate the infinitesimal action of the
Unitary group;
they are the moment maps on the coadjoint  orbit. The
dynamical system with
one of these as hamiltonian (energy) is not very
interesting,
since it
corresponds upon quantization to a free fermion system (see
below). The
hamiltonians of interest are quadratic functions of the
$\Phi$:
\beq
     H(\Phi)=[-\tr h\Phi+\half \tr \Phi\hat G (\Phi)]
\eeq
(The constant factor is chosen for later convenience).
Here $h$ is some constant hermitean matrix. $\hat G$ is a
positive
symmetric linear operator  on the vector space of Hermitean
matrices. In
terms of components,
\beq
     (\hat G \Phi)^i_j=G^{ik}_{jl}\Phi^l_k
\eeq
with
\beq
     G^{ik}_{jl}=G^{ki}_{lj}=G^{*jl}{ik}
\eeq
and $G^{ik}_{jl}\xi^{*j}_i\xi^l_k\geq 0$.  If
$G^{ik}_{jl}=B^i_j\delta^k_l+\delta^i_jB^k_l$, the quadratic
function will
be a constant on each connected component (depends only on
the trace of
$\Phi$). Similarly, if
$G^{ik}_{jl}=C^i_l\delta^k_j+\delta^i_lC^k_j$ it
will also be a constant. Therefore the tensor $\hat G$ can
be
chosen to be
traceless in all the indices at the cost  of changing the
hamiltonian by a
constant, which will not affect the equations of motion.
($h$ can also be
chosen traceless). The pair of irreducible tensors $h$ and
$G$
defines  a
`quadratic function' on the Grassmannian. Quadratic
functions on
Grassmannians have been studied before, \[guest] but the
ones we are
interested in do not seem to have a natural geometrical
meaning. Any $G$
can be expanded in terms of its eigenvectors,
$ G^{ij}_{kl}=\sum_a \gamma_a \xi^{i}_{ak}\xi^{*j}_{al}$,
$\gamma_a$ being
positive numbers. The particular tensors $\hat G$ that arise
in our theory
have the Weyl matrices as eigenvectors. It would be
interesting to obtain
an algebraic or charecterization of these tensors,
independent
of the
connection to fermionic systems.

 The extrema of the Hamiltonian function  (static solutions
of the
equations of motion) on the Grassmannians are of special
interest. A
variation that preserves the constraint $\Phi^2=1$ is of the
form
$\delta\Phi=[\Phi,U]$, where $U$ is an arbitrary anti--
hermitean matrix. Thus the condition for an extremum is
\beq
     [h-\hat G(\Phi),\Phi]=0.
\eeq
If $\hat G=0$, the solutions are of the form
$\Phi=\sum_{i}\eps_i
u_i\otimes u^{\dag i}$, with $\eps_i=\pm 1$ and $u_i$ being
the
eigenvectors of $h$. In each connected component, the
minimum is
corresponds to the choice
$\eps_i=-1(+1)$ for the lowest $m$ eigenvalues of $h$.
Even the case $\hat G=0$ corresponds to a highly nonlinear
classical
system, since the phase space is a curved manifold. The
minimum in
$Gr_{m,M}$  has the physical meanining of  the ground state
of a  free
fermionic system, with $m$ fermions.
The operator $N=\half(1-\Phi)$ (which satisfies $N^2=N$)
has as its
eigenvalues the occupation numbers of the fermionic system.
The additional
(quadratic) term represents interactions between the
fermions; this can of
course, distort the ground state.  The configuration
$\Phi_{m}$ with the
minimum energy
in each connected component $Gr_{m,M}$  is the semi-classical
approximation
to the ground state of the fermionic system in the expansion
we will introduce soon.

The time evolution of our system will be determined by the
ordinary
differential equations
\beq
    {i\over 2} {d\Phi\over dt}=[h-\hat G(\Phi),\Phi].
\eeq
Small oscillations around the minimum $\Phi_0$ in $Gr_{m,M}$
are of
physical interest; they correspond to charge density waves
(mesons) of the
fermionic system. (A small fluctuation will grow
exponetially
in time if we
expand around a critical point that is not a minimum.) The
equation for
small fluctuations is \beq
   {i\over 2}  {d\delta\Phi\over dt}=
          [h-\hat G(\Phi_0),\delta\Phi]+[\Phi_0,\hat
G(\delta\Phi)].
\eeq
In other words, the small oscillations satisfy
${d\delta\Phi\over dt}=\hat
K(\delta\Phi)$, $K$ being the linear operator on the
tangent space of the
Grassmannian at $\Phi_0$,
\beq
     \hat K=\;{\rm ad}\;(h-\hat G(\Phi_0))+\;{\rm ad}\;
\Phi_0\hat G.
\eeq

\subsect.{\it Action of the Grassmmanian system}

By a straightforward application of the usual point of view of
 on multivalued actions, we can construct an
action principle
that defines the classical system on the Grassmannian completely.

 In hamiltonian  mechanics, the symplectic form is usually
expressed in
canonical coordinates, $\omega=dq^i\wedge dp_i$. Given a
path in phase
space $c(t)=(q^i(t),p_i(t)$, its action is defined to be
\beq
     S=\int_c p_idq^i -\int_c H(p(t),q(t))dt
\eeq
the  subscript indicating that integrals are  evaluated
along the path $c$.
The path along which this is stationary is the classical
trajectory. If
the curve $c$ is closed, this can also be expressed as
\beq
     S=\int_s \omega -\int_{c=\pdr s} H(p(t),q(t))dt
\eeq
where $s$ is a surface in  the phase space whose boundary is
$c$. (We will
be interested only in situations where  such a curve
exists).
The latter
point of view is  suited to  situations where the symplectic
form is closed
but not exact.

This gives us an action principle on the Grassmannian
defining our system:
\beq
     S=-{i\over 8}\int \tr
\tilde\Phi(s,t)\d\tilde\Phi/dt\d\tilde\Phi/ds
dt\wedge ds-\int \tr[-h\Phi(t)+\half\Phi(t)\hat
G(\Phi(t))]dt.
\eeq
Here $\Phi(t)$ is a periodic function of $t$ decsribing a
closed curve on
the Grassmannian. Each component of the Grassmannian is
simply connected,
and $\tilde\Phi(s,t)$ is a deformation of this curve to a
point (i.e., it
is a surface whose boundary is $\Phi(t)$). Since $\omega$ is
closed, a
continuous deformation of the surface itself will not change
the value of
the action. However, if it is changed by a closed surface,
the value of the
action can jump by a constant. Since $\omega$ has been normalized such that its
integral over any closed 2--surface is an integer  multiple of $2\pi$, the
change of $S$ will also  be an integer multiple of $2\pi$.
 This has no effect on the
classical theory,
but does affect the quantum theory. In order that the quantum path integral
\beq
	\int {\cal D}\Phi e^{-{i\over \hbar} S}
\eeq
have a single valued integrand, ${1\over \hbar}$ must be an integer. This is
the integer denoted by $N_c$ below. We will see that the same restriction
arises in the canonical quantization method.

 \sect.{\it Quantization on the Grassmannian}

We have been studying a classical system whose phase space
is the
Grassmannian with the hamiltonian
\beq
     H=\tr[-h\Phi+\half \Phi\hat G(\Phi)].
\eeq
We will now show that the quantization of this system yields
a system of
fermions with  a hamiltonian that is quartic in the fermion
operators. In
this sense, the Grassmannian system we described earlier is
the classical
limit of a fermionic system. The classical limit of the
fermionic system
corresponds to the one in which the occupation numbers  go
to infinity.

There are several different routes to constructing the
quantum
theory. The
simplest is the algebraic (or canonical) quantization.

In the algebraic point of view, the classical system is
defined by the
(i) Poisson  brackets of a complete set of observables
\beq
     \{\Phi^i_j,\Phi^k_l\}=(\Phi^i_l\delta^k_j-
\Phi^k_j\de^i_l),
\eeq
(ii) the constraints satisfied by them,
\beq
     \Phi^i_j\Phi^j_k=\de^i_k
\eeq
and (iii) the hamiltonian
\beq
     H=\tr[-h\Phi+\half\Phi\hat G(\Phi)].
\eeq

One way to pass to the quantum theory is to associate to
each classical
observable $f$ (function on the phase space) a self--
adjoint
operator
$\hat f$ on a quantum Hilbert space such that
\beq
     [\hat f_1,\hat f_2]=i\hbar \widehat{\{f_1,f_2\}}.
\eeq
In conventional canonical quantization, one thinks of the
canonical
coordinates as  the basic observables and require them to
satisfy the
Heisenberg relations. For more  complicated operators, it is
possible to
preserve the connection of Poisson brackets to commutators
only upto higher
order terms in $\hbar$. The real parameter $\hbar$ measures
how much the
quantum theory deviates from the classical theory.

It is not a good idea to express our theory in terms of
canonical coordinates,  as such coordinates would be singular
somewhere in the phase
space. Instead, we will take the  Poisson brackets of the
$\Phi^i_j$ as the
analogues of the canonical commutation relations.
Quantization then
consists of finding a representation of the above algebra in
terms of
operators $\hat\Phi^i_j$
on a complex Hilbert space satisfying
\beq
     [\hat\Phi^i_j,\hat\Phi^k_l]=-
{i\hbar}(\hat \Phi^i_l\delta^k_j-
\hat \Phi^k_j\de^i_l).
\eeq
This means that the operators ${\hat\Phi\over \hbar}$
provide a
representation of the unitary Lie algebra. We can assume
this
representation is irreducible since otherwise, the whole
theory will split
up into a sum of irreducible components that do not interact
with each
other.  The real and imaginary parts of $\Phi^i_j$ are
observables, so that
the corresponding operators must satisfy the hermiticity
condition
\beq
     \hat\Phi^{\dag i}_j=\hat \Phi^j_i
\eeq
analogous to the classical condition $\Phi^{i*}_j=\Phi^i_j$.
This condition
requires the representation of the Lie algebra to be
unitary.

 Unitary irreducible representations of the Lie algebra
$\underline{U(M)}$
are well known. A basis in any irreducible representation is
given by the
simultaneous eigenvectors of ${1\over \hbar}\Phi^i_i$ ( no
sum
on $i$),
\beq
     {1\over \hbar}\Phi^i_i|w>=w_i|w>.
\eeq
In a unitary representation, these weights $w_i$ are
integers.
There is a
unique lowest weight vector, satisfying
\beq
     \hat\Phi^i_j|w>=0\for i>j
\eeq
 Also the lowest weight is  a nondecreasing sequence
$w_1\leq
w_2\cdots
w_M$. There is in fact a one--one correspondence between
such
sequences and
unitary irreducible representations of $U(M)$.

At this point our quantization scheme is ambiguous since
there
are many
irreducible representations. (The usual canonical
quantization is more
unique since the Heisenberg algebra has only one unitary
irreducible
representation).  Now recall the constraints
$\Phi^i_j\Phi^j_k=\de^i_k;\Phi^i_i=M-2m$ of the classical
theory. If we
require that these be true at least in the sense of an
expectation value on
the lowest weight state,
\beq
     <w|\hat\Phi^i_j|w><w|\hat\Phi^j_k|w>=\de^i_k;\quad
<w|\Phi^i_i|w>=M-2m
\eeq
we will get
\beq
     \hbar^2 w_i^2=1\quad \hbar \sum_i w_i=M-2m.
\eeq
This implies first of all that $\hbar$ is  the inverse of an
integer,
$\hbar={1\over N_c}$.( This integer is the number of
`colors'; the reason for this terminology will become clear
in a minute). Furthermore, it determines the lowest
weight vector
in terms of this integer:
\beq
     w_1=-N_c\cdots w_m=-N_c,w_{m+1}=N_c\cdots\eeq
Thus each connected component of the phase space corresponds
to an
irreducible representation of the unitary Lie algebra. The
representation is
unique upto the choice of the integer $N_c$, whose inverse
has the meaning
of $\hbar$. In particular, the large $N_c$ limit is the
classical limit of the theory.

  This representation of $\un{U(M)}$ can be decsribed more
explicitly in
terms of fermionic operators. Define a set of operators
satisfying the
anti--commutation  relations
\beq
     [a^{ i a},a^{\dag}_{jb}]_+=\de^i_j\de^a_b,\quad
[a^{ i a},a_{jb}]_+=0,\quad
 [a^{\dag i a},a^{\dag}_{jb}]_+=0
\eeq
where $i,j=1,\cdots M$, and  $a,b=1,\cdots N_c$. Then, there
is a representation of
this algebra on a fermionic Fock space ${\cal F}$ of
dimension
$2^{MN_c}$.
The bilinears
\beq
     \hat\Phi^{i}_{j}={1\over N_c}\sum
          [a^{ i a},a^{\dag}_{ ja}]
\eeq
provide a representation of our algebra of $\Phi$'s on this
Fock space.
However, as it stands this representation  is reducible. For
example,  the operators
\beq
     Q^a_b=-{1\over 2}[a^{ ia},a^{\dag}_{ib}]+{1\over
2M}\de^a_b[a^{ jb},a^{\dag}_{jb}]
\eeq
which generate a $SU(N_c)$ algebra commute with the
$\Phi^i_j$. To get an irreducible representation, we must
look at the subspace of vectors annihilated by these
operators.

 This $SU(N_c)$ symmetry is called `color' symmetry, since it
has to do with the number of copies of  otherwise identical
fermions. It has no meaning within our theory, and it is
natural that all our  states    be invariant under color. We
emphasize that $Q^a_b$  does  not generate an ordinary
symmetry that
commutes with just the hamiltonian; it is a {\it gauge}
symmetry,
that commutes with {\it all} the observables of our theory.

The fermion  number operator $a^{\dag}_{ia}a^{ia}$ also
commutes
with the $\Phi^i_j$ and so must be fixed to get an
irreducible representation.
If we restrict ourselves to the
states of zero  `color' and of fermion number $mN_c$,
\beq
     {\cal F}_{m0}=\{|\psi>\in {\cal F}| Q^a_b|\psi>=0;
 {1\over N_c} a^{\dag}_{ia}a^{ia}|\psi>=m|>\}
\eeq
we get an irreducible representation. The lowest weight
state is the state in which the states labelled by
$i=1,\cdots m$ are occupied, while the remaining ones are
unoccupied
\beq
          a^{\dag}_{ia}|w>=0\for i\leq m;\quad
a_{ia}|w>=0\for  i> m.
\eeq
The lowest weight can be calculated:
\beq
     \Phi^i_i|w>={1\over N_c}\sum_a[a^{\dag
ia},a_{ia}]|w>=\mp|w>\for i\leq m \;\;{\rm or}\;\; i>m
\eeq
agreeing with the earlier result. Thus the color singlet
fermionic  states do indeed provide  the particular
irreducible representation we need.

We can now show that the constraints  are true as operator
equations, in the Hilbert space ${\cal F}_{m0}$:
\beq
     \hat\Phi^i_j\hat\Phi^j_k=\delta^i_k,
\quad \tr\Phi^i_i=M-2m.
\eeq
The linear constraint is trivial to see. As for the
quadratic
constraint, if we  expand
\beqs{
     \hat\Phi^i_j\hat\Phi^j_k &=
{1\over
 N_c^2}\{[a^{ia},a^{\dag}_{ja}][a^{jb},a^{\dag}_{kb}]\}\cr
}\eeqs
it will be clear that every term can be written as a
bilinear
or as a bilinear multiplied by  $a^{\dag}_{ja}a^{jb}$ on the
left or
the right. Now we can use the condition of color invariance
to
reduce the latter terms also to bilinears. Thus we see that
\beq
     \hat\Phi^i_j\hat\Phi^j_k=a\delta^i_k+b\hat\Phi^i_k
\eeq
where $a,b$ are real numbers that may depend on $m$. We can
determine them them by taking the matrix elements of the
equation in the lowest weight state. If we put $i=k\leq m$,
the expectation value of the R.H.S. is $a-b$. For the
L.H.S.,
\beqs{
     {1\over
N_c^2}<w|[a^{ia},a^{\dag}_{ja}][a^{jb},a^{\dag}_{ib}]|w>=&
{1\over N_c^2}\sum_{j\leq
m}<w|a^{\dag}_{ja}a^{ia}a^{\dag}_{ib}a^{jb}|w>\cr
&={1\over N_c^2}||a^{\dag}_{ib}a^{ib}|w>||^2={N_c^2\over
N_c^2}=1\cr
}\eeqs
so that $a-b=1$. If we consider $i=k>m$, we will get
$a+b=1$.
Thus $a=1$ and $b=0$, proving our identity.

Thus we have  a representation of the commutation relations
that satisfies the constraint. To complete the construction
of
the quantum system, we  find the hamiltonian operator,
\beq
     \hat H=-h^i_j\hat\Phi^j_i+\half
G^{ij}_{kl}\hat\Phi^k_i\hat\Phi^l_j.
\eeq
This describes a system of fermions interacting through a
two
body `potential'. $G^{ij}_{kl}$ is the scattering amplitude
in
Born approximation for  a pair of fermions. When the
interaction is strong, it becomes very difficult to solve
the
system in this language. It is simpler to solve it in the
${1\over N_c}$ expansion, which corresponds to the semi--
classical approximation of the Grassmannian model.

The quantum fluctuations in the observables (real and
imaginary parts of $\hat
\Phi^i_j$) go to zero in the large $N_c$ limit. This is
clear
from the fact
that the commutators of the $\hat \Phi^i_j$  vanish as
${1\over N_c}$ and the
Heisenberg uncertainty relation
\beq
(\Delta A)^2(\Delta B)^2\geq {1\over 4}|<\psi|[A,B]|\psi>|^2
\eeq
for  a pair of hermitean operators. For finite $N_c$, one
can consider the
states in ${\cal F}_{m0}$ that minimize the uncertainty of
measuring the
various components of $\hat\Phi^i_j$. These would be the
analogue of minimum
uncertainty wave packets of quantum mechanics (`coherent
states'). One
invariant  measure \[perelomov] of this uncertainty is
\beq
     \Delta C_2=<\psi|\hat
\Phi^i_j\Phi^j_i|\psi>-
<\psi|\Phi^i_j|\psi><\psi|\Phi^j_i|\psi>.
\eeq
The states that minimize this are the lowest weight states.
i.e.,  staisfy
$\Phi^i_j|\psi>=0$ for some choice of basis in the Lie
algebra. In fact the set
of minimum uncertainty states is just the orbit of a
particular  lowest weight
state  under the action of $U(M)$. The isotropy group of
alowest weight state
is $U(m)\times U(M-m)$ so that this orbit is a Grassmannian.
We have here an
embedding of the Grassmannian into the Hilbert space ${\cal
F}_{m0}$. This
extends to a holomorphic embedding of $Gr(M)$ to the
projective space ${\cal
P}({\cal F}_{m0})$. The points on this submanifold of ${\cal
F}_{m,0}$ form an
overcomplete basis; i.e., any state can be written as a
linear
combination of
the points on the orbit of the lowest weight state.
For $N_c=1$ ( when the condition of color
invariance is
trivial) this is the standard Plucker embeddingof the
Grassmannian into the
projective space ${\cal P}(C^{{M\choose m}})$.

To summarize, the quantum Hilbert space of the dynamical
system on the
Grassmannian is the color singlet sector of a fermionic
system. ${1\over N_c}$
plays the role of $\hbar$ in this quantization. A
hamiltonian
on $Gr(M)$ that
is quadratic in $\Phi$ leads to a fermionic system with four
point (two--body)
interactions. The points of the Grassmannian correspond to
coherent states in
the Hilbert space; in the large $N_c$ limit, the quantum
fluctuations
dissappaear and we recover the classical model on the
Grassmannian.

 \sect.{ Hadrons in two dimensions}

\subsect.{The Grassmannian $Gr_1$}

We will first give a reasonably precise definition of the
phase space of  out theory, which will be an infinite rank
Grassmannian\[pressegal]. There is a homotopy invariant (called
`virtual rank') that plays an important role (baryon
number) in our theory. Without a precise definition it would
not be possible to see that there is such an invariant. We dont need to
maintain this level of mathematical rigor for other parts of the paper, since
the answers do ot depend critically on topological subtleties. It might be
useful to consult a standard reference on operator ideals (e.g. \[simon])
before reading this section; the summary given below is very brief and not very
pedagogical.

 Let $\Hi$ be a Hilbert space and $\Hi_\pm$ a pair of
infinite dimesnional
orthogonal subspaces such that $\Hi=\Hi_-\oplus \Hi_+$.
Define the
self--adjoint operator $\eps$ which has eigenvalues $\pm 1$
on $\Hi_\pm$. The
`restricted Unitary group'  $U_1$ is  defined by
\beq
     U_1=\{g|g^{\dag}g=1; [\eps, g] \in {\cal I}_2\}
\eeq
${\cal I}_2$ being the space of Hilbert--Schmidt operators.
(An operator $A$
is Hilbert--Schmidt if $A^{\dag}A$ is trace class: $\tr
A^{\dag}A<\infty$).
Thus the elements of $U_1$ are unitary and do not mix the
subspaces $\Hi_\pm$
`too much'. This group is of interest in quantum field
theory
\[pressegal]  because
the bilinears of a  $1+1$ dimensional fermionic field theory
form a
representation of its Lie algebra.  There, $\Hi$  is
the one particle Hilbert space; $\Hi_\pm$ are the positive
(negative)
eigenspaces of the Dirac hamiltonian. Thus $\eps$ is the
sign
of the first
quantized energy operator.
 In general, we would split the one particle Hilbert
space into $\Hi_-$, the states  with energy less than the
Fermi energy and
$\Hi_+$, the states with more than Fermi energy. In
nonrelativistic quantum
mechanics the dimension of $\Hi_-$
is finite; in relativistic quantum mechanics, this
space is infinite dimesnional.

We will be interested in the infinite rank Grassmannian
\beq
     Gr_1=U_1(\Hi)/U(\Hi_-)\times U(\Hi_+).
\eeq
$U(\Hi_\pm)$ is the subgroup of operators that leave the
negative (positive)
energy states invariant. (Equivalently, $U(\Hi_-)\times
U(\Hi_+)$ is the
subgroup that commutes with $\eps$).It can be shown that
this is an infinite
dimensional
manifold modelled on the Hilbert space ${\cal I}_2(\Hi_-
,\Hi_+)$ of
Hilbert--Schmidt  operators from $\Hi_-$  to $\Hi_+$. We
will
now define two
dimensional hadronic theory as a classical dynamical system
with $Gr_1$ as the
phase space.

 As before the
Grassmannian can be parametrized by operators $\Phi$:
\beq
     Gr_1=\{\Phi|\Phi^{\dag}=\Phi; \Phi^2=1;[\eps,\Phi]\in
{\cal I}_2\}.
\eeq
It is clear that any such operator can be reduced to $\eps$
by a unitary
transformation: $\Phi=g\eps g^{\dag}$. (The convergence
condition on $\Phi$
will ensure that $g$ itself satisfies the convergence
condition of
$U_1$.) But $g$ is ambiguous upto a right multiplication by
$h\in
U(\Hi_-)\times U(\Hi_+)$. Hence there is a 1-1
correspondence
between $\Phi$
and points of the coset space $Gr_1$. The point $\Phi=\eps$
will be the
`vacuum' configuration of the theory so in fact  it is more
convenient to
introduce the `normal ordered' variable $M=\Phi-\eps$. Thus,
we parametrize the
Grassmannian by
\beq
     Gr_1=\{ M|M^{\dag}=M; M^2+\eps M+M\eps=0; \tr
|[\eps,M]|^2<\infty\}.
\eeq

If we decompose $M$ into a $2\times 2$ matrix with respect to
the splitting
$\Hi=\Hi_-\oplus \Hi_+$ $M=\pmatrix{\alpha&\beta\cr
\beta^{\dag}&\delta}$,  the
convergence condition just says that $\beta$ is Hilbert--
Schmidt. The quadratic
constraint now implies  that $\alpha$ and $\delta$ are
trace--
class. To see
this write $M=g\eps g^{\dag}-\eps$ where, $g=\pmatrix{a&b\cr
c&d\cr}$.  Here
$b,c$ are Hilbert--Schmidt and $a$ and
$d$ are only bounded.
 Then,
$\alpha=-aa^{\dag}+bb^{\dag}+1$ etc.Using the unitarity of
$g$, ($
aa^{\dag}+cc^{\dag}=1$), we see that $\alpha=
bb^{\dag}+cc^{\dag}$ which is trace class. This technical
remark will
be useful in
defing the topological invariant called `virtual rank'
below.

The group $U_1$ has an infinite number of connnected
components labelled by an
integer. To see this, consider $g=\pmatrix{a&b\cr c&d\cr}$
decomposed into
blocks as before; $a:\Hi_-\to \Hi_-,b:\Hi_+\to \Hi_-,c:\Hi_-
\to
\Hi_+,d:\Hi_+\to \Hi_+$.
Since $g$ is invertible and $b,c$ compact,it follows that,
$a$ and $d$ are
Fredholm operators
of opposite index.

We digress a little to give some basic definitions and
results
of operator theory. There are many stanadard references
where
results are stated more precisely e.g., \[simon].
 An operator is compact if there is a
sequence
of finite rank
 operators that converges to it in the operator norm.
Hilbert--Schmidt and
trace class operators are in particular compact. Not all
bounded operators are
compact: for example the identity is not. An
operator
is Fredholm if it
is invertible modulo the addition of a compact operator. The
kernel of an
operator $a$ is the subspace of all $|\psi>$  such that
$a|\psi>=0$. The co--kernel
is the set of $|\psi>$ such that $a^{\dag}|\psi>=0$. The
dimension of the
kernel is {\it not} a continuous function of $a$. However,
the
index of $a$
which is the dimension of its kernel minus that of its
cokernel is invariant
under continuous deformation of $a$.

Returning to our context, the index of the submatrix $a$ of
$g$  is the net number of states in $\Hi_-$
which are pushed
out
to $\Hi_+$ by $g$. The index of $a$ is invariant under a
continuous
deformation of $g$.  The index of $d$ is just minus that of
$a$: as we change $g$, if a certain number of
states cross from
$\Hi_-$ to $\Hi_+$, an equal number will cross in the
opposite
direction. Two elements of $U_1$ with the same index are
connected  by  a continuous path: $U_1$ is the disjoint
union
of connected
components labelled by the index.

It is easy to give an example of a unitary operator $g$ with
non--
zero index for
$a$. Suppose we  label an orthonormal basis of $\Hi$ by
integers,  with $e_n$
for $n\leq 0$ ($n>0$) spanning $\Hi_-$($\Hi_+$). An element
of
the connected
component of $U_1$ with index $k$ is the shift operator
$\sigma_k$
\beq
     \sigma_k e_n=e_{n+k}.
\eeq
Note that if we had not imposed the condition that
$[\eps,g]$
is
Hilbert--Schmidt, the Unitary group would have been
contractible; there would
be no such topological invariant. We will see that this
topological invariant
has the meaning of baryon number in our hadronic theory,
which
is an essential
feature. More generally $\pi_{2m}(U_1)=Z$ for even homotopy
groups and
$\pi_{2m+1}(U_1)=0$ for odd homotopy groups.

The Grassmannian is also a disjoint union of connected
components labelled by
an integer.  This was already true for finite rank
Grassmannians. In fact
$-\half\tr(\Phi-1)$ is the dimension of the subspace in
which
$\Phi$ has   eigenvalue $-1$ ( the rank of the
Grassmannian).
For our present case this would diverge. We can get a
convergent quantity by
subtracting the rank of the vacuum $\Phi=\eps$; i.e.,
consider
$-\half\tr(\Phi-\eps)=-\half\tr M$.
This is the `virtual rank' of the Grassmannian, the
difference
between the
number of negative eigenvalues of $\Phi$ and those of
$\eps$.
(A technical
remark: we can define $\tr M$ to be
$\tr \alpha+\tr\delta$. Here
$M=\pmatrix{\alpha&\beta\cr\beta^{\dag}&\delta}$
in a basis in which $\eps=\pmatrix{-1&0\cr0&1\cr}$. We
showed
earlier that
$\alpha$ and $\delta$ are trace class. This indirect
definition of  $\tr M$ is necessary since $M$ itself is not
trace class.)  In fact if $M=g\eps
g^{\dag}-\eps$, it
is easy to
see that $-\half\tr M=\;{\rm index}\;(a)$. (For example, if
$g=\sigma_1$ the
shift operator, $(e_n,\sigma_1\eps \sigma_1-\eps
e_n)=\sgn(n-1)-\sgn(n)$. Then
$\tr M=-2$ and the index of $a$ is 1.)

It is best to think of $\Hi$ as the space $L^2(R;C)$ of
complex valued
functions on the real line. In this parametrization,
$\eps$ is the  sign of the momentum operator,
\beq
     \eps=\int [dk]e^{ikx} \sgn(k).
\eeq
Here,
\beq
     [dk]={dk\over 2\pi}.
\eeq
We can also describe it by its integral kernel in position
space
\beq
     \eps(x,y)={i\over 2\pi}{\cal P}\left(1\over x-y\right);
\eeq
This  is, $i$ times the  well--known Hilbert transform operator. The
Grassmannian
can
be parametrized
by the integral kernel $M(x,y)$ of the operator $M$. The
condition of
self--adjointness is just
\beq
 M^*(x,y)=M(y,x).
\eeq
The quadratic constraint is  the nonlinear integral equation
\beq
     \int dy[
M(x,y)M(y,z)+\eps(x,y)M(y,z)+M(x,y)\eps(y,z)]=0.
\eeq
The convergence condition is best understood in terms of the
Fourier transform
\beq
     \tilde M(p,q)=\int e^{ipx-iqy}M(x,y)dxdy.
\eeq
The condition is that the off--diagonal component of $\tilde
M$ with $p$ and
$q$ having opposite sign has finite norm:
\beq
     \int_{p>0} [dp]\int_{q<0}[dq] |\tilde M(p,q)|^2
<\infty.
\eeq
The topological invariant (`virtual rank') of the
Grassmannian is then
\beq
     w(M)=-\half\int dx M(x,x)
\eeq
which we will be shown to have the physical meaning   baryon
number.

\subsect.{Symplectic structure and Poisson Brackets}

We saw  in the finite rank case, that the Grassmannian is a
symplectic manifold
with $\omega=-{i\over 8}\tr \Phi(d\Phi)^2$. The analogue in
infinite dimensions
is  just $\omega=-{i \over 8}\tr (M+\eps)(dM)^2$. A tangent
vector at the point
$M=0$  is a matrix of the form $U=\pmatrix{0&u\cr
u^{\dag}&0\cr}$ with $\tr
u^{\dag}u<\infty$. It is straightforward to see that
$\omega(U,V)=-{i\over 4}\tr(u^{\dag}v-v^{\dag}u)$. This
shows
that $\omega$ exists
and is non--degenerate at the vacuum.
That $\omega$ exists and is
non--degenerate at an arbitrary  point follows from its
homogenity (see
below). The proof
that $d\omega=0$ goes through as before.

The transformation  $\Phi\to g\Phi g^{\dag}$ or $M\to
gMg^{\dag}+g[\eps,g^{\dag}]$ leaves the symplectic form
invariant. From general
principles, there must be functions that generate the
corresponding infinitesimal  canonical
transformations,$V_u=i[u,\eps+M]$.
(The matrix
$u=\pmatrix{a&b\cr b^{\dag} &d}$  must have diagonal
elements$a,b$ bounded and
off--diagonal elements $c,d\in {\cal I}_2$  in order that
$e^{iu}$ be in $U_1$.) That is, there must be a function
$f_u$
such that  $\omega(V_u,.)=df_u$. In the finite dimensional
case, this is just $f_u(\Phi)=\tr u\Phi$.  In the finite
dimensional case, these are just linear functions of $\Phi$.
For our present case, again $\tr u\Phi$ will diverge; we
must
subtract the
vacuum contribuition and consider instead $ f_u(M)=-\half\tr
uM=-\half\tr
u(\Phi-\eps)$. (It is easy to check that the trace in
$f_u(M)$
does indeed converge.)  But we will pay a price for this
vacuum subtraction. The Poisson brackets of the $f_u$ will
no
longer be that of the Lie algebra of $U_1$: there will be a
non--trivial
central extension.

The  computation that verifies that $f_u(M)$ generates $V_u$
is straightforward:
\beqs{
     \omega(V_u,.)&=- {i\over 8}\tr (\eps+M)[ V_u(M),dM]=
{-i\over 8}\tr [V_u,\eps+M]dM\cr
&={1\over
8}\tr[\eps+M,[\eps+M,u]]dM=-\half\tr udM=df_u.\cr
}\eeqs
Now we get
\beq
     \{f_u,f_v\}(M)=-\Lie_{V_u}f_v(M)=-
i\tr[u,\eps+M]v=f_{(-i)[u,v]}(M)+{i\over 2}\tr [\eps,u]v.
\eeq
This is the central extension of the Lie algebra of $U_1$.
We
cannot remove
the central term by redefining $f_u(M)\to f_u(M)+\tr\eps u$
since the second
term does not converge in general. This nontrivial extension
of the unitary Lie
algebra was found in a different context by Kac and
Petersen.
In the co--ordinate basis, these Poisson brackets are,
\beq
{i\over 2}\{M(x,y),M(z,u)\}=\delta(y-z)[M(x,u)+\eps(x-u)]-\delta(x-
u)[M(z,y)+\eps(z-y))].
\eeq\label{winfty}
This may be verified by multiplying both sides by
$u(y,x)v(u,z)$ and integrating over $x,y,z,u$ to get the
ealier Poisson bracket relation.

Occasionally the above Lie algebra  is also called the
$W_\infty$ algebra.
This algebra arises
in the matrix model approach to string theory. Our theory of
hadrons
is a nonlinear matrix model, the dynamical variable being
$M(x,y)$. We believe
that in fact our theory is a string field theory in two
target
space dimensions
written in light cone gauge. Finding an ungauge fixed
(manifestly
reparametrization invariant form) of this theory is an
extremely interesting
problem. It is already known that the linear approximation
to
our theory (which is described by 't Hooft's integral equation) is
equivalent to the  free
Nambu string in two target space dimensions \[bars]

The Poisson brackets above along with the constraints
can be thought of as the definition of our phase space. The
interpretation in
terms of the Grassmannian shows how natural these Poisson
brackets are, given
the constraints.
Conversely, given the above Poisson brackets, we can look
for
homogenous
symplectic manifolds on which the Lie  group acts. These are
given by the
co--adjoint orbits, of which  the Grassmannian is one of the
simplest. The
other co--adjoint orbits do not describe fermion fields upon
quantization.

\subsect.{The Hamiltonian}

We will now describe two dimensional classical hadron
theory.
The most
convenient co--ordinate system of Minkowsky space   for our
purpose is the one
with metric $ds^2=du(du+2dx)$. In terms of conventional
space-
-time
co--ordinates $ds^2=(dx^0)^2-(dx^1)^2$ and $u=x^0-
x^1,x=x^1$.
The surface $u=$ constant is a null line and  $x$ is a co--
ordinate in this
null direction. The vector $\d/du$ (for fixed $x$) is just
the same as the
time--like vector $\d/d{x^0}$. Thus momentum component $p_u$
is just energy
while the $x$ component of momentum (which we will call
just
$p$) is
$p=p_0+p_1$. The mass shell condition for a particle of mass
$\mu$ is
\beq
     2p_up-p^2=\mu^2\implies p_u=\half[p+{\mu^2\over p}].
\eeq
The Lorentz transformation is, in this co--ordinate system,
\beq
     u\to e^{\theta}u\quad x\to -\sinh\theta u+e^{-\theta} x
\eeq
where $\theta$ is the rapidity. From the invariance of
$p_udu+pdx$, we get
\beq
     p\to e^{\theta}p\quad p_u\to e^{-\theta}p_u+\sinh
\theta
p.
\eeq
The phase space will be given by the initial conditions on
the
surface $u=0$;
the dynamics will be given by the hamiltonian function $H$
which will generate
translation
in $u$. There should also be functions  $L$ and $P$
generating
Lorentz
tranformations and translations, satisfying the Poincare
algebra:
\beq
     [P,L]=0\quad [L, P]=P\quad  [L,H]=-H+P.
\eeq

A dynamical system is completely specified by the phase
space
(a manifold
along with a symplectic form) and a Hamiltonian. We will
define two dimensional
hadron theory by choosing $(Gr_1,\omega)$ as the phase
space.
Thus our dynamical
variable is a function of two points $M(x,y)$ satisfying the
constraint
\beq
 M^*(x,y)=M(y,x),\quad
     \int dy[
M(x,y)M(y,z)+\eps(x,y)M(y,z)+M(x,y)\eps(y,z)]=0.
\eeq
This specifies the initial condition at $u=0$. The value at
any later vaue of
$u$ is given by solving the classical equations of motion
\beq
     \d{M(x,y)}/du =\{H,M(x,y)\}.
\eeq
We choose the  hamiltonian to be
\beq
     H=\int dxdy h(x-y)M(x,y)-\half \tilde g^2\int dxdy G(x-
y)M(x,y)M(y,x).
\eeq
The operators $h$ and $G$ are defined by
\beq
     h=\half(p+{\mu^2\over p}); \quad G={1\over p^2},
\eeq
where $p=-i{d\over dx}$. The theory depends on the
 parameters $\tilde g$ and $\mu$  with
the dimensions of mass. We will show later that $\mu$ is
related to the quark
mass and $\tilde g$ to
the gauge coupling constant of 2DQCD.
The  integral kernels of the operator  are
\beq
     h(x-y)=\half [-i\delta'(x-y)+{i\over 2}\sgn(x-y)]
,\quad G(x-y)=-\half|x|.
\eeq
The choice of $h$ as the kinetic energy operator is
motivated
by the earlier
discussion of the mass shell condition.

Although the field variable is bilocal, the theory we have
just defined is a
Poincare invariant field theory. Translation invariance in
the
$x$ direction is
obvious.  Under Lorentz tranformations, the field transforms
as
\beq
     M(x,y)\to e^{\theta} M(e^{\theta}x,e^{\theta}y).
\eeq
It is straightforward to check that $H$ and $P=\tr pM$
transform as the
components of a Lorentz vector under this  transformation.
Once we assume the
form \(ham) for the hamiltonian, specific expressions for
$h$
and $G$ are
required by Lorentz invariance alone. It should be possible
to
show by an
argument based on string theory that \(ham) is the only
possible form for the
hamiltonian.

The equations of motion that follow from this hamiltonian
are,
\beqs{
     {i\over 2}\d{M(x,y)}/du&={i\over 2}\{H,M(x,y)\}=\int dz [ h(x-z)M(z,y)-
M(x,z) h(z,y)]\cr
&+\tilde g^2\int dz [G(y-z)\eps(x,z)M(z,y)-G(z-
x)\eps(z,y)M(x,z)]\cr
          &+\tilde g^2\int dz M(x,z)M(z,y)[G(y-z) -G(z-
x)].\cr
}\eeqs\label{eqmotion}
By construction, the quadratic constraints are preserved
under
time evolution.

The solution $M=0$ is the vacuum solution. Small
oscillations
around this are
described by the linear approximation to the above equation.
These will
describe mesons. There are also stationary solutions whaich
are very far from
this vacuum solution, which describe baryons. The quantity
$B=-\half\tr M$, is
an integer valued topological invariant and hence is
conserved
by time
evolution. The lowest energy solution with $B=1$ should be
identified with the
baryon. Linearizing \(eqmotion)  around this stationary
solution  will describe
meson--baryon scattering; the bound states in this
linearized
equation will
dsecribe excited states of the baryon ( analogues of
$\Delta,N^*$ etc.).

\subsect.{ Linear Approximation}

Let us now study the linearization of the above theory
around
the vacuum $M=0$.
We must linearize not only the equations of motion but also
the constraints.
If we drop the second order term, the constraint becomes,
\beq
     \int dy [\eps(x,y)M(y,z)+M(x,y)\eps(y,z)]=0.
\eeq
Since our equations are translation invariant, it is better
to
use the Fourier
transformed variable
\beq
     \tilde M(p,q)=\int dxdye^{i(-px+qy)}M(x,y).
\eeq
The constraint becomes:
\beq
     [\sgn(p)+\sgn(q)]\tilde M(p,q)=0.
\eeq
In addition the hermiticity consition becomes
\beq
     \tilde M^*(p,q)=M(q,p).
\eeq
This means that our field operator has only off-diagonal
components; i.e.,
those with opposite signs for $p$ and $q$. (Of course, this
is just the
statement that $M$ is tangential to $Gr_1$ at the  origin).
The translation,
$M(x,y)\to M(x+a,y+a)$, and
\beq
     \tilde M(p,q)\to e^{i(-qa+pa)}\tilde M(p,q).
\eeq
Thus $P=p-q$ has the meaning of total momentum. Due to the
hermiticity
condition,it is sufficient to consider the case $p>0$ and
$q<0$, so that the
total momentum is positive. Define now the dimensionless
variable $\xi={p/P}$;
then $\xi$ varies in the range $0\leq \xi\leq 1$. We can use
$P$ and $\xi$ as
our independent variables. Set
\beq
     \chi(P,x)= P \tilde M(\xi P,(1-\xi)P).
\eeq

The linearized equations of motion is, in position space,
\beqs{
     i\d{M(x,y)}/du&=\int dz [ h(x-z)M(z,y)-M(x,z)
h(z,y)]\cr
&+\tilde g^2\int dz [G(y-z)\eps(x,z)M(z,y)-G(z-
x)\eps(z,y)M(x,z)].\cr
}\eeqs\label{leqmotion}
In momentum space this becomes,
\beq
     i\d{\tilde M(p,q)}/du=   [h(p)-h(q)]\tilde
M(p,q)+\tilde
g^2[\sgn(p)-\sgn(q)]\int
{[dr]\over r^2} \tilde M(p-r,q-r).
\eeq
It is clearly natural to assume that the $u$ dependence is
exponential $$\tilde M(p,q,u)=e^{-iP_u u}\tilde M(p,q)$$. In
terms of the variable $\chi(P,\xi)$ we get the eigenvalue
equation
\beq
     M^2\chi=[{\mu^2\over \xi}+{\mu^2\over 1-
\xi}]\chi(\xi)+4\tilde g^2\int_0^1{d\xi'\over (\xi-
\xi')^2}\chi(\xi')
\eeq
where $M^2=2P_uP-P^2$ is the mass$^2$ of the meson. This is
precisely 't Hooft's integral equation. He has shown that
all
solutions are bound states ( discrete eigenvalues) and that
spectrum is assymptotically $M_n^2\sim n\tilde g^2.$

\subsect.{ The baryon solution}

If ours is to be a complete theory of hadrons, it must
contain
baryons as well as mesons. From Witten's \[witten] arguments
one should expect baryons to arise in a Hartree
approximation.
On the other hand, in the Skyrme  model, they arise as
solitons in a theory whose small oscillations are mesons.
Normally, this meson theory is only known approximately.
However, we have an {\it exact} theory of hadrons in two
dimensions. This gives us an oppurtunity to study the baryon
within this exact theory. Also, we will get a description of
a
topological soliton in a bilaocal field theory. Since we
should expect our bilocal theory to be a string field
theory,
we will also  get a glimse of how a topological soliton
looks like in string theory.

 We already showed that there is a topologically  conserved
quantum number $w(M)=-\half\tr M$ in our theory. It takes
integer values. (It will be shown later that this indeed
corresponds to baryon number of  2DQCD.) We will look for a
configuration that minimizes the energy, and with $w(M)=1$.
Unlike the vacuum solution, it will have non--zero energy:
it
is a topological soliton. Upon quantization, its mass will
seen to be $O(N_c)$, since ${1\over N_c}$ has the meaning of
$\hbar$ in the quantum theory.  We will see that the
equation
satisfied by  soliton field configuration  has a natural
interpretation in terms of a Hartree approximation for the
quark wavefunction.

Our first task is to produce an ansatz that has virtual rank
one. It will be much more convenient to deal with functions
of
one variable rather two, so we will seek a separable ansatz
of
the form
\beq
     M_1(x,y)=-2\psi(x)\psi^*(y).
\eeq
In order to have virtual rank one  we must require the
function $\psi$ to be normalized:
\beq
     \int dx|\psi(x)|^2=1.
\eeq
Now let us see the condition on $\psi$ implied by the
quadratic constraint, $M_1^2+\eps M_1+M_1\eps=0$. It is
clear
that $\psi(x)\psi^*(y)$ is the kernel of a projection
operator, so that $M_1^2=-2M_1$. Thus we get $\eps M_1 +
M_1\eps=2M_1$. In terms of $\psi$ this is just the condition
$\eps\psi=\psi$. That is, $\psi$ is an element of $\Hi_+$,
the
space with eigenvalue $+1$ for $\eps$:
\beq
     \int \eps(x-y)\psi(y) dy=\psi(x).
\eeq

 Thus our ansatz $M_1$ is indeed an element of $Gr_1$ with
virtual rank one.( It is trivial to check that $[\eps,M]$ is
a
rank one operator, so that the convergence condition $\tr
|[\eps,M]|^2<\infty$ is satisfied).

In fact $M_1$ is unchanged under the change of phase
$\psi(x)\to e^{i\theta}\psi(x)$. The ansatz gives a map from
the projective space ${\cal P}(\Hi_+)$ to $Gr_1$. We will
see
that $\Hi_-$ is the set of negative energy states for the
quarks, which are occupied in the ground state of 2DQCD. To
produce a baryon, we must put $N_c$ quarks in one positive
energy state as well. In the ground state, all these
quarks
will have the same wavefunction, apart from a color factor which is completely
anti--symmetric. This is the meaning of
$\psi$ in the quark model. Note that  the condition that
$\psi$ is a positive energy state arises naturally in the above
formalism. It will be crucial for the stability of the
soliton.

The energy of this configuration is obtained by substituiting
the ansatz into the hamiltonian \(ham). The first term in
the
hamiltonian becomes $\tr hM=(\psi,h\psi)$ or, $\int
\psi^*(x)h(x-y)\psi(y)dxdy$. Noting also that
$M(x,y)M(y,x)=|\psi(x)|^2|\psi(y)|^2$  we get
\beq
     H(\psi)=\int \psi^*(x)h(x-y)\psi(y)dx dy-\half \tilde
g^2\int G(x-y)|\psi(x)|^2\psi(y)|^2 dxdy.
\eeq
It is now clear that the first term is just the kinetic
energy
of the quark wavefunction. The second term is the potential
energy due to a Coulomb field generated by the quarks. There
is an attractive linear potential between the quarks. This
is
exactly what we would have obtained in a Hartree
approximation. There are $N_c$ quarks in a baryon, whose
color
indices must anti--symmetrized to get a singlet state. The
quarks being fermions, the wavefunction must be symmetric in
the remaining ( position and flavor) indices. Thus
effectively
they are bosons and  the ground state wavefunction of all
 the quarks will be the  same. The second term in  the energy
then
describes the interaction with the mean field.

The condition $\eps\psi=\psi$ ensures that the kinetic
energy
$(\psi,h\psi)$ is positive. $h=\half[p+{\mu^2\over p}]$ is
not
in general positive; $\Hi_+$ is the susbspace in which it is
positive. If it had not been for the constraint that
$\psi\in
\Hi_+$, the energy  would not be bounded below and our
soliton
would have been unstable. The constraint is best understood
in
momentum space. If we define the Fourier transform
\beq
     \tilde \psi(p)=\int \psi(x) e^{-ipx} dx
\eeq
we see that it has support only for positive $p$. In terms
of
position space, $\psi(x)$ must be a function that can be
analytically continued to the  upper half of the complex
plane.

Thus we must minimize the energy over all configurations
$\tilde\psi(p)$ satisfying
\beq
     \tilde \psi(p)=0\;{\rm for}\; p<0\quad
\int_0^{\infty}|\psi(p)|^2 [dp]=1.
\eeq
The Kinetic energy is simple to understand in momentum space
while the potential energy is simpler in position space. The
energy is
\beq
     H(\tilde \psi)=
\half \int_0^{\infty} [p+{\mu^2\over p}]|\tilde
\psi(p)|^2[dp]
+\half\tilde g^2\int V(x)|\psi(x)|^2 dx.
\eeq
Here $V(x)$ is the solution to the diffrential equation
\beq
     V''(x)=|\psi(x)|^2
\eeq
with the boundary conditions $V(0)=0$ and $V'(0)=0$. This
variational problem can be reduced to a nonlinear integral
eigenvalue problem.
We have not been able to solve this problem analytically.
The problem was solved by numerically minimizing the energy
in
Ref.\[rajeevetal].

 Here we will find the qualitative behaviour of the soliton
using a simple     variational approximation. An ansatz that
satisfies the positivity of
energy and the
normalization condition is $\tilde\psi(p)=Npe^{-pa}$, where
$a>0$ is a variational parameter.(The even simpler
possibility $\tilde\psi(p)=Ne^{-pa}$ has to discarded
because
it has infinite potential energy). Then, $\psi(x)={N\over
2\pi}{i\over (x+ia)^2}$.  Note that this has a double pole
on
the lower half plane, but is analytic in the upper half
plane.
The normalization condition is ${N^2\over 2\pi}=4a^3$. The
kinetic energy integral is easily done:
\beq
     \half\int [p+{\mu^2\over
p}]|\tilde\psi(p)|^2[dp]=\half[{3\over 2a}+\mu^2a].
\eeq
The differential equation
\beq
     V''(x)=|\psi(x)|^2={2a^3\over \pi}{1\over (x^2+a^2)^2}
\eeq
has solution
\beq
     V(x)={1\over\pi} x\arctan{x\over a}.
\eeq
Note that $V(x)\sim \half|x|$ as $|x|\to \infty$, as
expected.
Thus the potential energy becomes
\beq
     \half\tilde g^2\int V(x)|\psi(x)|^2 dx=
\half I\tilde g^2 a
\eeq
where
\beq
     I={4\over \pi}\int_0^{\infty}{x\arctan x\over 1+x^2}
dx=2.
\eeq
Thus we estimate the energy of the soliton to be
\beq
     H(a)=\half[{3\over 2a}+{\mu_{\rm eff}^2\over a}]
\eeq
where the quark mass $\mu$ is replaced by the effective
mass:
\beq
     \mu_{\rm eff}^2=\mu^2+2\tilde g^2.
\eeq
As the mass of the quark goes to zero, the lightest meson
also
has a very small mass \[thooft]. But we see that even as the
mass of the quark goes to zero, the mass of the baryon
remains
finite! The effective mass of the quark in the baryon is
heavier by an amount proportional to the gauge coupling
constant. That is, the `constituent quark mass' is of the
order of the gauge coupling constant, even as the current
quark mass goes to zero. We have just established a
phenomenon
analogous to chiral symmetry breaking in two dimensions. (Of
course the chiral symmetry itself cannot break spontaneously
in two dimensions; the meson of small mass also has small
coupling, so that it is not a Goldstone boson.)

In the quantum theory, there is an overall factor of $N_c={1\over \hbar}$ in
front of the action ( and hence the hamiltonian) so that the mass of the baryon
is $N_c$ times the minimum of $H$ as determined above. Thus our estimate
$\mu_{eff}$ is the mass of the baryon divided by $N_c$, which can be thought of
as the constituent quark mass. In the case of mesons, the energy of the vacuum
configurations are zero so that the masses of the mesons come from small
oscillations around the vacuum. These are therefore $\hbar={1\over N_c}$ order
smaller than the baryon mass, hence are  $O(N_c^0)$.

\sect.{ Quantization of Hadron Theory}

As in the finite dimensional case, we can find an action principle for QHD. The
quantum path integral would be
\beq
	\int {\cal D}\Phi e^{-{i\over \hbar}S}
\eeq
where
\beq
S=\int \omega({\pdr \tilde M\over dt},{\pdr \tilde M\over ds}dt\wedge ds-\int
H(M(t))dt.
\eeq
Again, $\tilde M$ is an extension of $M$ to a function of two variables $t,s$
such that the boundary value is the closed curve $M(t)$. As before,
$H^2(Gr_1)=Z$ and  the integral of $\omega$ on a generator of $H^2(Gr_1)$ is
$2\pi$. Thus ${1\over \hbar}=N_c$ must be an integer in order that
 $e^{-{i\over \hbar}S}$ be single valued. Since at the moment it is difficult
to define such path integrals rigorously, we will follow instead a canonical
(algebraic) point of view. The definition of the path integral can be
accomplished by expoliting localization formulae.

We can now quantize the hadron theory by algebraic(canonical)
methods. We would
 convert the Poisson brackets into commutation relations:
\beq
     [\hat f_u,\hat f_v]={i \hbar}\big(\hat f_{-i[u,v]}+\tr
[\eps,u]v\big).
\eeq
Equivalently, we look for operator--valued distribuitions
$\hat M(x,y)$ satisfying,
\beqs{
     [\hat M(x,y),\hat M(z,u)]&=-i\hbar\big( \delta(y-z)
[\hat M(x,u)+\eps(x,u)]-\cr
&\delta(u-x)[\hat M(z,y)+\eps(z,y)]\big).\cr
}\eeqs
These will provide a  a  representation for the Lie algebra
of
$U_1$ on some space ${\cal F}$. (Each matrix element $\hat
M(x,y)$ of $M$ must itself be an operator on ${\cal F}$.)

 The representation we pick must be unitary:
\beq
     \hat M(x,y)^{\dag}=\hat M(y,x)
\eeq
and irreducible. (If the $\hat M(x,y)$ are to be a complete
set of observables, the only operators that commute with
them
must be multiples of the identity; but then the
representation
is irreducible.)
Also the quadratic constraint must be satisfied at least
upto
terms that vanish as $\hbar \to 0$.

Furthermore, the hamiltonian must become a well--defined
(self--adjoint) operator that is bounded below. Even for  the
simplest case of the hamiltonian with $\tilde g=0$ this
imposes a nontrivial constraint on the representation.
It is best to discuss this case in momentum  picture, so we
define
$\htd M(p,q)=\int dxdy\hat M(x,y)e^{-ipx+qy}.$
 The commutation relations become,
\beqs{
[\htd  M(p,q),\htd  M(r,s)]&={-i\hbar}\big[2\pi
\delta(q-r)[\hat M(p,s)+\sgn(p)\delta(p-q)]\cr
&- 2\pi \delta(s-
p)[\hat M(r,q)+\sgn(r) \delta(r-s)]\big].\cr
}\eeqs
It is easy to recognize this as a representation of the
central extension of of the unitary Lie algebra, if we write
this in terms of ${\htd M\over \hbar}$:
\beqs{
[{\htd  M(p,q)\over \hbar} ,{\htd  M(r,s)\over \hbar}]&=-
i\big[2\pi
\delta(q-r)[{\hat M(p,s)\over \hbar}+{1\over
\hbar}\sgn(p)\delta(p-q)]\cr
&-2\pi \delta(s-
p)[{\hat M(r,q)\over \hbar}+{1\over \hbar}\sgn(r) \delta(r-
s)]\big].\cr
}\eeqs
Note that the central terms are proportional to ${1\over
\hbar}$.

The hamiltonian is
$\hat
H_0=\int [dp]h(p) \hat{\tilde  M}(p,p)$
 In the
case of interest to us, the operator $h$ has an infinite
number of  eigenvectors with a negative eigenvalue: elements
of $\Hi_-$. It is clear that acting with $\htd M(p,q)$ on
any
state
will add $p-q$ to its momentum. By operating with $\htd
M(p,q)$ with $p<q$ repeatedly,  we may be able to construct
states of arbitarily negative
momentum (and hence energy). The only way to avoid this
catastrophe is to eventually arrive at  a state
(`vacuum') $|0>\in {\cal F}$ such that
\beq
     \hat M(p,q)|0>=0. \for p<q
\eeq
 It is clear that $\htd M(p,q)$ for $p<q$
can be thought of as  the negative roots of the Lie algebra
of
$U_1$: we have just shown that the representation of ${\cal
F}$ must be a lowest weight representation. If the
representation is to be irreducible, the lowest weight
vector
must be unique. Also,  every vector in ${\cal F}$ can be
written as linear combination of the vectors $\htd
M(p_1,q_1\cdots \htd M(p_r,q_r)|0>$ obtained from the
vacuum by positive roots:$p_1>q_1\cdots p_r>q_r$. The vacuum
itself is an eigenvector of the diagonal generators:
\beq
      {\htd M(p,p)\over \hbar}|0>= w(p)|p>
\eeq
(To make the theory mathematically rigorous, we can suppose
that
space is  a circle so that the momenta take on just discrete
values. At the end the radius of this circle can be taken to
infinity.)
Lowest weight unitary representations are classified by the
eigenvalues of the diagonal elements (lowest weight)\[kp].
If this representation of the Lie algebra can be
exponentiated
to one of the Lie group, the weights $ w(p)$ must be
integers.
   Moreover,   $1\over \hbar $ is an integer $N_c$ for the
central term to be exponentiated.
(In fact the condition that there be a positive invariant
inner product already requires $w(p),N_c$ to be integers).
The quantization of $N_c$  is analogous to the quantization
of
the level number in
the classification of lowest weight unitary representations
of
the affine Kac--Moody algebras. Indeed, we saw that $1\over
\hbar$ appears only as a
coefficient of the central term of the Lie algebra.

Now we are interested in representations in which the
quadratic constraint on $M$ are preserved in some sense. As
in
the finite dimensional case, we could simply impose the
quadratic condition on  the expectation value of $\htd M$ on
the lowest weight state. This will lead to the condition
\beq
     [\hbar w(p)+\sgn(p)]^2=1.
\eeq
If we restrict to a topological sector where the baryon
number
is $B$ we have the linear constraint,
\beq
     -\half\sum_p<w|\htd M(p,p)|w>=B\implies \sum_p[\hbar
w(p)]=B.
\eeq
These conditions uniquely determine the lowest weight $w(p)$
in terms of $N_c$ and $B$:
\beq
     w(p)=N_c[\sgn(p-p_F)-\sgn(p)]
\eeq
the `Fermi momentum' $p_F$ being fixed to be $p_F=B$.

This particular representation of the unitary Lie algebra
can
be written in terms of fermionic variables, just as we did
in
the finite dimensional case.
Introduce the fermionic fields satsfying the canonical anti-
commutation relations:
\beq
     [\t \chi^a(x),\t
\chi^{\dag}_b(x)]_+=\delta^a_b\delta(x-
y),\quad [\chi^a(x),\chi^b(y)]_+=0,\quad
[\chi^{\dag}_a(x),\chi^{\dag}_b(y)]_+=0.
\eeq
Here $a,b=1\cdots N_c$ is the `color' index. Then, we define
the lowest weight state ( vacuum of the free field theory)
by
\beq
     \t \chi^a(p)|0>=0,\for p>p_F,\quad \t
\chi^{\dag}_a(p)|0>=0\for p\leq p_F,
\eeq
where $\hat \chi$ is the Fourier transform of $\chi$. We can
now verify that
\beq
     \hat M(x,y)={1\over N_c}:\chi^{\dag}_a(x)\chi^a(y) :
\eeq
satisfies the commutation relations above. The central terms
arise from the fact that $\hat M$ has to be  normal ordered
with respect to the above lowest weight state.

This representation of the unitary Lie algebra on the
fermionic Fock space   is  reducible; in order to get an
irreducible representation, we must impose the conditions
\beq
     Q_a^b|\psi>=0
\eeq
on all the physical states, $Q_a^b=\int[:\chi^{\dag}_a(x)\chi^a(x)-{1\over
N_c}\delta^a_b\chi^{\dag}_c\chi^c(x)]dx$ being the `color
charge'. On each subspace of color singlet states of fixed
baryon number, our algebra has an irreducible
representation.In terms of the fermion fields, baryon number
is
\beq
     B={1\over N_c}\int:\chi^{{\dag}}_a(x)\chi^a(x):dx.
\eeq

Now we can express the hamiltonian in terms of $\hat M$ and
hence $\chi$.
\beq
     H=\int dxdy h(x-y)\hat M(x,y)-\half \tilde g^2\int dxdy
G(x-y)\hat M(x,y)\hat M(y,x).
\eeq\label{ham}
As before, the operators $h$ and $G$ are defined by
\beq
     h=\half(p+{\mu^2\over p}); \quad G={1\over p^2},
\eeq
where $p=-i{d\over dx}$. We have constructed the
representation and the normal ordering of $\hat M$ such that
the first term is finite. The interaction term involving
$G(x-y)$ is also now finite.  (If the coupling constant had been
dimensionless as in the Thirring model, this would have been
true only after a coupling constant renormalization.)

 Next we will show that the above quantization of
the hadron thoery is equivalent to two dimensional QCD. More
precisely, we will show that it is equivalent to the color
singlet sector of QCD. As part of this we will see that
$N_c$
corresponds to the number of colors and $B$ to the baryon
number of QCD.

We will find it convenient to formulate QCD in the light--
cone
type co--ordinates defined earlier:
\beq
     ds^2=du(du+2dx),\quad u=x^0-x^1,x=x^1.
\eeq
The action of 2DQCD is
\beq
     S=\int \big\{-{1\over 4}F^a_{\mu\nu b}F^{\mu\nu b}_a+
\bar q_a[\gamma^\mu(-i\delta^a_b\pdr_\mu-gA_{\mu b}^a)-
m\delta^a_b]q^b\big\}d^2x
\eeq
We are considering the theory with just one flavor and
$SU(N_c)$ colors. Thus $A^a_{\mu b} $ are traceless
hermitean
matrices, with  $a,b=1\cdots N_c$.  The generalization to
several flavors is quite straightforward. The variables $q,\bar q$ are to  be
viewed as anti--commuting (Grassmann--valued) in a path integral approach.

The Dirac matrices satisfy the relations
\beq
     (\gamma^u)^2=0,\quad (\gamma^x)^2=-1,\quad
[\gamma^u,\gamma^x]_+=2.
\eeq
A convenient representation is
\beq
     \gamma^u=\pmatrix{0&0\cr 2&0\cr},\quad
\gamma^x=\pmatrix{0&1\cr -1&0\cr}.
\eeq
It will will be convenient label the componets of the quark
fields as follows,
\beq
     q={1\over \surd 2}\pmatrix{\chi\cr \eta}.
\eeq
Then,
\beq
     \bar q=q^{\dag}\gamma^0={1\over \surd
2}\pmatrix{\eta^{\dag}&\chi^{\dag}\cr}.
\eeq
As for the gauge fields, it is convenient to choose the
light--cone gauge, $A_x=0$  and denote the remaing component by
$A_u=A$. With these choices,  the action of the theory
becomes,
\beq
     S=\int \big\{-\half\tr(\pdr_x A)^2+\chi^{\dag}[-
i\pdr_u-
gA]\chi+\half[-i\eta\pdr_x\eta+i\chi^{\dag}\pdr_x\chi]+\half
m[\chi^{\dag}\eta+\eta^{\dag}\chi]\big\}dxdu.
\eeq
Now it is clear that the fields $\eta$ and $A$ do not
propagate; there equations of motion can be used to
eliminate
them in favor of $\chi$:
\beq
     -i\pdr_x\eta+m\chi=0,\quad \pdr_x^2
A^a_b+g\chi^{\dag}_b\chi^a=0.
\eeq
After this elimination, the action becomes
\beq
     S=-i\int\chi^{\dag}_a\pdr_u\chi^adxdu-
\int\chi^{\dag}_ah\chi^a dxdu+{g^2\over 2}\int G(x-
y)\rho^a_b(x)\rho^b_a(y)dxdydu.
\eeq
Here
\beq
	\rho^a_b(x)=\chi^{\dag}_b(x)\chi^a(x)-{1\over
N_c}\delta^a_b\chi^{\dag}_c(x)\chi^c(x)
\eeq
is the color charge density. Also, $h$ and $G$ are defined as before. We
already begin to see the sort of expressions that appear in QHD.

The first term in the action  implies that in canonical quantization, $\chi$
and $\chi^{\dag}$ are conjugate; the remaining terms determine the hamiltonian.
The commutation relations are
\beq
     [\t \chi^a(x),\t
\chi^{\dag}_b(x)]_+=\delta^a_b\delta(x-
y),\quad [\chi^a(x),\chi^b(y)]_+=0,\quad
[\chi^{\dag}_a(x),\chi^{\dag}_b(y)]_+=0.
\eeq
 The, Fock representation of these relations  can now be constructed,
based on the naive vacuum state
\beq
     \t \chi^a(p)|0>=0,\for p>0,\quad \t
\chi^{\dag}_a(p)|0>=0\for p\leq 0.
\eeq
The hamiltonian is
\beq
	H=\int:\chi^{\dag}_ah\chi^a :dx+{g^2\over 2}\int G(x-
y):\rho^a_b(x)\rho^b_a(y):dxdy.
\eeq
A normal ordering with respect to the above vacuum is necessary to make the
hamiltonian of the quantum theory well-defined.
It is now possible to write the hamiltonian in terms of the color singlet
operator
\beq
	\hat M(x,y)={1\over N_c}:\chi^{\dag}_a(x)\chi^a(y):.
\eeq
We have already seen that this operator satisfies the commutation relations of
QHD. Thus we see the operator $\hat M(x,y)$ is a quark--antiquark bilinear. We
see that $N_c$ is the number of colors and $B=\int \hat M(x,x)$  the baryon
number of QCD.

The kinetic energy term is already of the form $\tr hM$. The interaction term
also can be written in terms of $\hat M$ if we use the Fierz identity
\[rajeevetal]. This will lead to precisely the hamiltonian of QHD. Since the
calculation has been done elsewhere \[trieste],\[rajeevetal],\[kikkawa], we
will not carry it
out  here.  Thus we see that the theory we have been studying is
equivalent to the color singlet sector of QCD. The parameters $\mu,\tilde g$
are related to the quark mass $m$ and gauge coupling constant $g$ by the
formulae,
\beq
	\mu^2=m^2-{g^2N_c\over 4\pi},\quad \tilde g^2=g^2 N_c.
\eeq

It is to be noted that although the parameter $m$ in the 2DQCD lagrangian is
traditionally called the quark mass, the quark is not asymptotic  particle in
the theory and has no well--defined mass. If the scattering of color singlet
particles produce only color singlet
particles, the S--matrix of QHD will be unitary. This would amount to a proof
of confinement of QCD. It seems likely that such a proof can be made to all
orders in the $1\over N_c$ expansion.

{\bf Acknowledgement}

The author thanks A. P. Balachandran, P. Bedaque, G. Ferretti, K. Gupta, S.
Guruswamy, I. Horvath and  T. Turgut for discussions at various stages in the
preparation of this paper. The hospitality of the Institute for Advanced Study
during the Spring of 1991, where much of this work was done, is also
acknowledged. This work was supported in part by the US Department of Energy,
 Grant No. DE-FG02-91ER40685.

\vfill\eject

 \sectnumber=0

\def\Lie{\;{\cal L}}
\def\implies{\Rightarrow}
\def\for{\;\;{\rm for}\;\;}
\def\implies{\Rightarrow}
\def\for{\;\;{\rm for}\;\;}
\def\un#1{underline{#1}}
\def\Hi{{\cal H}}
\def\Fm{{\cal F}_{0m}}

\centerline{\it Appendix: Hartree--Fock Theory and Grassmannians}

\sect.{Slater Determinants and Grassmannians}

In this appendix we will show the analogue of our approach to  2DQCD, in
electronic physics is just the Hartree--Fock approximation. It is hoped that
our point of view in terms of Grassmannians to Hartree--Fock theory is of some
use in atomic and condensed matter physics.

Let ${\cal H}=L^2(R^3,C^2)$  be the one particle Hilbert space of electrons; we
will think of its elements as complex valued functions $u(x,\sigma)$ of
position  $x\in
R^3$ and  spin $\sigma=\pm 1$. \footnote{$^*$}{Often it will be convenient to
suppress the spin variable, and let $x$ stand for $(x,\sigma)$, $\int dx$ for
$\sum_{\sigma}\int dx$ etc. The norm, for example, is then
$||u||^2=(u,u)=\int dx u^*(x)u(x)$. } The Hilbert space of an $m$-electron
system is the exterior power $\Lambda^m(\Hi)$. A wavefunction in this space is
a completely antisymmetric function $\psi(x_1,\cdots, x_n)$.
The hamiltonian of the $m$ electron system is  the following operator on
$\Lambda^m(\Hi)$:
\beq
	H=\sum_1^m[-\nabla_a^2+V(x_a)]+\sum_{1\leq a<b\leq m} G(x_a-x_b).
\eeq
For an atom, $V(x)={-Z\over |x|}$ describing the Coulomb attraction to the
nucleus and $G(x-y)={1\over |x-y|}$. (We are using units with ${\hbar^2\over
2m}=1$ and $e=1$. ) The ground  state is the minimum of
the expectation value of the hamiltonian over all states of unit norm:
\beq
	E_0=\inf_{||\psi||=1}(\psi,H\psi).
\eeq
If $||\psi||=1$, the expectation value of $H$ can be written as
\beq
	<\psi|H|\psi>=\tr h\rho_1+\tr G_2\rho_2
\eeq
where $h=-\nabla^2+V$ and the `one--particle density matrix' $\rho_1$  is  a
selfadjoint operator $\Hi$ with kernel
\beq
	\rho_1(x,y)=m\int
\psi^*(x,x_2,\cdots x_m)\psi(y,x_2,\cdots,x_m)dx_2\cdots dx_m.
\eeq
Also $G_2$ and $\rho_2$ are selfadjoint operators on the two particle Hilbert
space $\Lambda^2(\Hi)$ with kernels
\beq
	\rho_2(x,y;z,u)={m\choose 2}\int
\psi^*(x,y,x_3,\cdots x_m)\psi(z,u,x_3,\cdots,x_m)dx_3\cdots dx_m
\eeq
\beq
	G_2(x,y;z,u)=\half[\delta(x-z)\delta(y-u)-\delta(x-u)\delta(y-z)]G(x-y).
\eeq
The one and two particle density matrices are normalized as below:
\beq
	\tr \rho_1=\int \rho_1(x,x)dx=m,\;\quad
\tr\rho_2=\int \rho_2(x,y;x,y)dxdy={m\choose 2}.
\eeq

The problem of finding the ground state of this system is very hard. The idea
of the Hartree--Fock theory is to approximate the wavefunction by a Slater
determinant
\beq
	\psi(x_1\cdots x_m)={1\over \surd m!}\left|\matrix{u_1(x_1)&\ldots&
u_1(x_m)\cr
       		\vdots&\ddots&\vdots\cr
                 u_m(x_1)&\ldots &u_m(x_m)\cr}\right|.
\eeq
The one particle wavefunctions $u_a\in \Hi$  are orthonormal
$(u_a,u_b)=\delta_{ab}$.
A change $u_a(x)\to g_{ab}u_b(x)$ will change the Slater determinant only by a
phase $\det g$, so it doesnt change the physical state represented by it.

The Hartree--Fock approximation to the ground state energy is the minimum over
all choices of $u$ of the expectation value
$E^{HF}(u)$  of the hamiltonian in this state:
\beq
	E^{HF}_0=\inf_{(u_a,u_b)=\delta_{ab}} E^{HF}(u).
\eeq
Clearly, $E_0\leq E^{HF}_0$.
$E^{HF}$  can be woked out to be:
\beqs{
	E^{HF}(u)&= \sum_{a=1}^m\int dx \int dx dyu^*_a(x)[-\nabla^2+V(x)]u_a(x)+\cr
&\sum_{a<b}\int dx dy G(x-y)
               {1\over \surd 2}[u^*_a(x)u^*_b(y)-u^*_b(x)u^*_a(y)]
	{1\over \surd 2}[u_a(x)u_b(y)-u_b(x)u_a(y)].\cr}\eeqs
However, it is somewhat unnatural to think of the $u_a(x)$ as the variational
parameters of Hartree--Fock theory; a change $u_a(x)\to g^a_b u_b(x)$ by a
unitary matrix does not affect the multiparticle state. In fact the larger $m$
is, the larger the number of such spurious degrees of freedom in $u_a$. In
quantum field theory, $m$ is infinity, this is particularly awkward.  We will
therefore develop a new point of view in terms of Grassmannians which has a
natural generalization to quantum field theory.
Perhaps this point of view is useful even in atomic physics.

To get a more  intrinsic form for $E^{HF}$, let us compute $\rho_1$ and
$\rho_2$ for this ansatz. We will get\beq
	\rho_1(x,y)=P(x,y);\quad\rho_2(x,y;z,u)=\half[P(x,z)P(y,u)-P(x,u)P(y,z)]
\eeq where
\beq
P(x,y)=\sum_{a=1}^m u^*_a(x)u_a(y).
\eeq
$P$ is the projection operator to the $m$--dimensional subspace spanned by
$u_a$.
The expression for $\rho_2$ in terms of $P$ can be written as
\beq
	\rho_2=P\wedge P
\eeq
where the wedge product of two operators $A,B$ on $\Hi$ is the operator on
$\Lambda^2(\Hi)$
\beq
	(A\wedge B)(u,v)=(Au)\wedge (Bv).
\eeq
Thus we get
\beq
	E^{HF}(P)=\tr hP+\tr G_2 P\wedge P.
\eeq
In coordinate basis,
 \beqs{
	E^{HF}(P)=&\int dx [\left({\pdr^2 P(x,y)\over \pdr x\pdr
y}\right)_{x=y}+V(x)P(x,x)]+\cr
&\half\int dxdy G(x-y)[P(x,x)P(y,y)-P(x,y)P(y,x)]\cr
}\eeqs
The first term represents the kinetic energy and the second term the potential
energy due to the nucleus. The third term is the energy of Coulomb repulsion of
the electrons (`direct  energy'). The last term is the `exhange energy' due to
the antisymmetry of the wavefunctions. (The fact that Hartree--Fock theory can
be expressed in this way is well known \[lieb]). The
Hartree--Fock ground state energy is the  minimum of $E^{HF}(P)$ over all
projection operators of rank $m$:
\beq
	E^{HF}_0=\inf_{P^2=P;\tr P=m} E^{HF}(P).
\eeq

The set of all projection of operators of finite rank defines an infinite
dimensional analogue of the Grassmannian\footnote{$^*$}{We can require $P$ to
be compact rather than finite rank; any compact projection operator is also
finite rank.}
\beq
	Gr(\Hi)=\{ P|P^{\dag}=P;P^2=P; P {\rm\ is\ finite\ rank} \}.
\eeq
Then $\tr P$ exists and is an integer; the Grassmannian is a union of connected
components labelled by the trace; each connected compoenent is the homogenous
space
\beq
	Gr_m(\Hi)=U(\Hi)/U(m)\times U(\Hi)
\eeq
just as  the finite dimensional case. $Gr_m(\Hi)$ is an infinite dimensional
manif
Each point  in the Grassmannian represents an $m$ dimensional subspace of
$\Hi$, the eigenspace of $P$ with eigenvalue 1. There is an embedding of
$Gr_m(\Hi)$ into ${\cal P}(\Lambda^m\Hi)$, ( generalization of the Plucker
embedding) which is precisely the meaning of the Slater determinant. To
construct this  analogue  of the Plucker embedding, pick an orthonormal basis
in the subspace of $P$:
\beq
	P=\sum_a u_a\otimes u^{\dag}_a
\eeq
and define the corresponding point in $\Lambda^m \Hi$ to be
\beq
	u_1\wedge u_2\wedge\cdots w_m.
\eeq
Again, this vector changes by the phase $\det g$ as we change the orthonormal
basis. The Slater determinant is precisely this wedge product written in the
position basis. There is a one--one correspondence between states  of the
Slater determinant type and points on the Grassmannian; there are no longer any
spurious variables in the problem. The ground state   just corresponds to the
minimum of $E^{HF}(P)$ on the Grassmannian.

This point of view  can be used perhaps to prove the existence of extrema (
critical points)  for the Hartree--Fock energy. We make a digresion to point
out a connection to Morse theory. $Gr_m(\Hi)$ is the well known model for the
Classifying Space \[milnorch] of $U(m)$. So its homotopy type is particularly
simple. Also, the cohomology of $Gr_m(\Hi)$ has a simple description;
any element can be written as a linear combination of wedge products of the
forms
\beq
	\omega_{2i}=\tr P(dP)^{2i}\;\;{\rm for}\;\; i=1,2\cdots\leq m.
\eeq
Furthermore there are no relations among these generators.(There would be some
relations among these generators if were considering instead the Grassmannian
of a finite dimensional vector space.) The odd cohomology groups vanish.
The number of generators of $H^{2k}$ is  equal to the number of ways $k$written
as the sum of the numbers $1,2,\cdots m$.
Thus the generating function of the Betti numbers (`Poincare polynomial') is
\beq
	P(t):=\sum_{p=0}^{\infty}{\rm dim} H^p t^p=\prod_{i=0}^m{1\over (1-t^{2i})}
\eeq
If $h$ and $G_2$  are reasonable operators on $\Hi$ and $\Lambda^2\Hi$,
${\cal E}$ will be a differentiable function on $Gr_m(\Hi)$. Then Morse theory
would guarantee the existence of extrema for ${\cal E}$. The minimum would be
the Hartree--Fock approximation to the ground state; the other extrema of
finite index will represent excited states. This approach is not  possible if
we think of the HF energy as a function of $u_a$.For, the space of such
orthormal frames on $\Hi$ (the Stiefel manifold of $\Hi$) is contractible;
Morse theory would be trivial. There should also be,  in general, a continuum
in the
case of atomic physics, corresponding to the scattering states.  One can
enclose the system in a large box so that all states are discrete and then the
energy function will only have isolated critical points.

The  weak Morse inequalities say that the number of critical points of index
$p$, $N_p$ is greater than or equal to ${\rm dim} H^p$. Thus  $N_{2k}$ is
greater than or equal to the number of ways in which $k$ can be written as a
sum of the numbers $\{1,\cdots,m\}$.  One can see that this agrees with the
counting based on the interpretation of the excited states of    an atom in
terms  of electron--hole pairs. The index is the number of directions in which
the energy of a state decreases, which is related to the number of states with
energy less than the given state.

\sect.{ Second Quantized approach to Hartree--Fock Theory}

It is clear that mean field theory is a sort of semi--classical approximation
to the atomic  system: the quantum fluctuation are small. So there must be some
sense in which Hartree--Fock theory describes a classical theory whose
quantization gives the atomic
hamiltonian. However this cannot be the
conventional classical theory of particles moving in a  Coulomb potential,
since the atom has no stable ground state in that approach. There must be some
other classical theory, with a stable ground state, whose quantition also
leads to the atomic hamiltoinian. We will show that this is a system whose
phase space is the Grassmannian, and that time--dependent Hartree--Fock theory
is equivalent to hamiltonian dynamics on this space.

To begin with let us  rewrite the problem in second quantized language.
Define the creation--annihilation operators associated to the Hilbert space
$\Hi$:
\beq
	[a^{\dag}(x),a(y)]_+=\delta(x-y);\quad [a^{\dag}(x),a^{\dag}(y)]_+=0;\quad
[a(x),a(y)]_+=0.
\eeq
There is a representation of this algebra on the Fock space ${\cal
F}=\oplus_{m=0}^{\infty}\Lambda^m(\Hi)$. The vacuum is the state containing no
electrons
\beq
	a(x)|0>=0.
\eeq
The number operator is
\beq
	N=\int a^{\dag}(x)a(x)dx
\eeq
 and its eigenspace with eigenvalue $m$ is $\Lambda^m(\Hi)$. The elements of
$\Lambda^m(\Hi)$ are of the form
\beq
	|\psi>=\int \psi(x_1,\cdots,x_m)a^{\dag}(x_1)\cdots a^{\dag}(x_m)dx_1\cdots
dx_m|0>.
\eeq
In this language, the hamiltonian  is the operator on ${\cal F}$
\beq
	\hat H=\int a^{\dag}(x)ha(x)dx	+\int dxdyG(x-y)a^{\dag}(x)a^{\dag}(y)a(y)a(x).
\eeq
The relation to the earlier language is,
\beq
	\hat H|\psi>=|H\psi>.
\eeq

Now suppose we consider a generalization of this  atomic physics problem where
each electron has another quantum number (`color') $\alpha=1,\cdots,N_c$. We
have the operators satisfying
\beq
	[a^{\dag \alpha}(x),a_\beta(y)]_+=\delta^\alpha_\beta\delta(x-y);\quad
[a^{\dag \alpha}(x),a^{\dag \beta}(y)]_+=0;\quad [a_\alpha(x),a_\beta(y)]_+=0.
\eeq
There is again a representation of this on the Fock space. But we will allow as
physical states only those annihilated by
\beq
	Q^\alpha_\beta=\int dx[a^{\dag \alpha}(x) a_\beta(x)-{1\over N_c}
a^{\dag \gamma}(x)a_\gamma(x)].
\eeq
These operators  generate an $SU(N_c)$ symmetry. Let us denote this `color
invariant'
subspace by ${\cal F}_0$:
\beq
	|\psi>\in {\cal F}_0\iff Q^\alpha_\beta|\psi>=0.
\eeq
Generalize the hamiltonian to the operator,
\beqs{
	\hat H=&{1\over N_c}\int a^{\dag \alpha}(x)ha_\alpha(x)dx	+
{1\over 2N_c^2}\int dxdyG(x-y)\cr
&[a^{\dag \alpha}(x)a^{\dag \beta}(y)a_\beta(y)a_\alpha(x)+a^{\dag
\alpha}(x)a^{\dag \beta}(y)a_\alpha(y)a_\beta(x)].\cr
}\eeqs
This is a generalization of the hamiltonian familiar from atomic physics.
It is clear that if $N_c=1$, the color singlet condition on the states becomes
trivial and the hamiltonian reduces to the previous one. This theory of colored
fermions is  of physical interest only for $N_c=1$. Yet, we will show  that
Hartree--Fock ( mean field) approximation is the large $N_c$ limit of this
theory. Expansion in powers of ${1\over N_c}$ will yield the expansions around
mean field theory. Although $N_c=1$ in the physical case, we know that this
${1\over N_c}$ expansion is in fact a good approximation method in atomic
physics. Thus we will see that the  conventional mean field theory can be
understood in terms of a `replica trick'.

To see this, introduce the operators
\beq
	\hat P(x,y)={1\over N_c}a^{\dag i}(x)a_i(y).
\eeq
They satisfy the commutation relations
\beq
[\hat P(x,y),\hat P(z,u)]={1\over N_c}[\delta(y-z)\hat P(x,u)-\delta(u-x)\hat
P(z,y)]
\eeq
of an infinite dimensional unitary Lie algebra. \footnote{$^*$}{Strictly
speaking we should regard the Lie algebra as consisting of the smeared objects
$\int dxdy\hat P(x,y)K(x,y)$  where $K$ is the Kernel of a {\it compact}
operator. Then there will be no divergence problems in the representation
theory of this algebra}
 The trace $\int \hat P(x,x) dx$ is  the number operator divieigenvalues of
this operator are positive integers; let ${\cal F}_{0m}$ be the
set of all states with eigenvalue $m$.

The operators $\hat P (x,y)$ forms a complete system of observables on $\Fm$.
i.e., any operator that commutes with $\hat P(x,y)$ for all $x,y$ is a multiple
of the identity. This follows from Schur's lemma if we can see that the
representation of the above Unitary algebra on $\Fm$ is irreducible. In fact
the irreducible representations of the  Lie algebra of compact selfadjoint
operators (and the corresponding group of unitary operators that differ from
the identity by a compact operator) has been worked out by Kirillov.
The representations are classified by Young tableaux, just as in the finite
dimensional representation theory. Now it is straightforward to verify that the
representation on $\Fm$ is the one with a rectangular Young diagram with $N_c$
columns and $m$ rows.

Although the $\hat P(x,y)$ form a complete set of observables on $\Fm$, they
are not all independent of each other; they satisfy a quadratic constraint. To
see this, first define the normal ordered product of two $\hat P$'s:
\beq
:\hat P(x,z)\hat P(y,u):=\hat P(x,z) \hat P(y,u)-{1\over N_c}\delta(y-z)\hat
P(x,u).
\eeq
In terms of the creation--annihilation operators, this means that all the
creation operators stand to the left of the annihilation operators:
\beq
:\hat P(x,z)\hat P(y,u):=-{1\over N_c^2}a^{\dag \alpha}(x)a^{\dag \beta}(y)
a_{\alpha}(z)a_{\beta}(u).
\eeq
The normal ordered product will have finite matrix elements in the limit $z\to
y$.

Between a pair of elements $|\psi>,|\psi'>$ of $\Fm$, these operators satisfy
the identity
\beq
	<\psi'|\int :\hat P(x,y) \hat P(y,u)  :dy |\psi>=<\psi'|\left(1-{m\over
N_c}\right)\hat P(x,u)|\psi>.
\eeq
The proof is to note that
\beqs{
	<\psi'|\int :\hat P(x,y) \hat P(y,u) dy :|\psi&=
{1\over N_c^2}\int <\psi'|a^{\dag \alpha}(x)
a^{\dag \beta}(y)a_{\alpha}(y)a_{\beta}(u)|\psi> dy\cr
&=-{1\over N_c^2}\int <\psi'|a^{\dag \alpha}(x)a_{ \beta}(u) a^{\dag
\beta}(y)a_{\alpha}(y)|\psi> dy\cr
&+{1\over N_c^2}\delta^{\beta}_\beta
<\psi'|a^{\dag \alpha}(x)a_\alpha(u)|\psi>\cr
&=\left(1-{m\over N_c}\right)<\psi'|\hat P(x,u)|\psi>.
}\eeqs
We have used
\beq
	\int dy a^{\dag \beta}(y)a_\alpha(y)|\psi>=-{1\over
N_c}\delta^\beta_\alpha\int dy a^{\dag\gamma}(y)a_\gamma(y)|\psi>=
m \delta^\beta_\alpha|\psi>
\eeq
for $\psi\in \Fm$.

Now we can  reformulate the theory entirely in terms of the color singlet
variables $\hat P$. The hamiltonian is
\beqs{
	\hat H=&\int dx [\left({\pdr^2 \hat P(x,y)\over \pdr x\pdr
y}\right)_{x=y}+V(x)\hat P(x,x)]+\cr
&\half\int dxdy G(x-y)[:\hat P(x,x)\hat P(y,y):-:\hat P(x,y)\hat P(y,x):].\cr
}\eeqs
The normal ordering is necessary to avoid  self--energy terms.

Now we see that with our definitions, $N_c$ appears only as a coefficient of
the commutation relations of the $\hat P(x,y)$. Thus ${1\over N_c}$ plays the
role of $\hbar$; it determines  the uncertainty in measuring the different
components of $\hat P(x,y)$ simultaneously. There are certain states in the
space $\Fm$ that minimize this uncertainty; they are the analogue of the
minimum uncertainty wavepackets (coherent states) of the usual canonical
quantization. An invariant measure of the uncertainty ( or the size of quantum
fluctuations) is \[perelomov]
\beq
\int[<\psi|:\hat P(x,y)\hat P(y,x):|\psi>-
  <\psi|\hat P(x,y)|\psi><\psi|\hat P(y,x)|\psi>]dxdy.
\eeq
The states that minimize are the highest weight states of some basis.
Explicitly, they are of the form
\beq
	|u>=\int dx_1\cdots dx_m u_1(x_1)\cdots u_m(x_m) B^{\dag}(x_1)\cdots
B^{\dag}(x_m)|0>
\eeq
where
\beq
	B^{\dag}(x)=a^{\dag 1}(x)\cdots a^{\dag N_c}(x).
\eeq
Clearly $B^{\dag}(x)$ createSince the quantum fluctuation around the
expectation values vanish in these
states, they will have a good classical limit.

 In the limit $N_c=1$ these are precisely the states represented by the Slater
determinant. More generally, they describe an embedding of the Grassmannian
$Gr_m(\Hi)$ into the projective space ${\cal P}(\Fm)$, generalizing the Plucker
embedding. We can now calculate the  expectation  value of $\hat P$:
\beq
	<u|\hat P(x,y)|u>=\sum_a u^*_a(x)u_a(y):=P(x,y).
\eeq
This is just the projection operator to the subspace spanned by the $u_a$.
Furthermore, the expectation value of the hamiltonian is
\beqs{
	<u|\hat H|u>=&\int dx [\left({\pdr^2 P(x,y)\over \pdr x\pdr
y}\right)_{x=y}+V(x)P(x,x)]+\cr
&\half\int dxdy G(x-y)[P(x,x)P(y,y)-P(x,y)P(y,x)].\cr
}\eeqs
Note that so far we have not made any approximations: these are the exact
expectation values. Since these Slater--type states are not the exact
eigenfunctions of $\hat H$, time evolution will take them into more complicated
states.

However, in the large $N_c$ ( semiclassical) limit the deviations will be small
and we will be able to decsribe the system completely in terms of the classical
vraiable $P(x,y)$. The classical phase space is the set of all projection
operators of rank $m$; i.e., the Grassmannian $Gr_m(\Hi)$. The commutation
relations of the variables $\hat P$ tend to Poisson brackets of the classical
variables:
\beq
	\{P(x,y),P(z,u)\}=i[\delta(y-z)P(x,u)-\delta(u,x)P(z,y)].
\eeq
These are precisely the ones that follow from the standard symplectic form on
the Grassmannian:
\beq
	\omega=-{i\over 8}\tr P (dP)^2
\eeq

Thus we see that  the limit $N_c\to \infty$ of our generalized atomic system is
a classical dynamical system with $Gr_m(\Hi)$ as phase space, $\omega$ as
symplectic form, and
\beqs{
E^{HF}(P)=&\int dx [\left({\pdr^2 P(x,y)\over \pdr x\pdr
y}\right)_{x=y}+V(x)P(x,x)]+\cr
&\half\int dxdy G(x-y)[P(x,x)P(y,y)-P(x,y)P(y,x)]\cr
}\eeqs
as the hamiltonian. In particular, the ground state is determined by the
minimum of $E^{HF}$. Also the excited states of the system correspond to
stationary points of higher index, as anticipated earlier.

Our derivation shows that time dependent Schrodinger equation tends in the
limit $N_c\to \infty$ to the equations
\beq
	{ d{P(x,y)}\over dt}=\{P(x,y),E^{HF}(P)\}
\eeq
as the equations of motion.  This could be an interesting approach to
calculating the scattering amplitudes of atoms by other atoms. The small
oscillations around a stationary point will have some charecterictic
frequencies. These will determine the correction to order ${1\over N_c}$ of the
energy levels. It is clearly possible to develop a systematic semiclassical
expansion in ${1\over N_c}$.

Thus we have a classical system whose quantization gives the atomic
hamiltonian. In fact, quantization can be performed by finding an irreducible
representation of the algebra above. The parameter $N_c$ ( effectively ${1\over
\hbar}$) must be a positive  integer  in order that a highest weight
representation  exist. The physically interesting
value happens to be $N_c=1$. The success of Hartree--Fock theory shows that
even in this extreme case, the ${1\over N_c}$ expansion gives reliable results.

{\bf References}\hfill

\def\ni{\noindent}

 \ni\strongstring.  D. J. Gross and W. Taylor Nucl.Phys. B403,395 (1993).

\ni\berezin. F. A. Berezin, Comm. Math. Phys. 63, 131 (1978).

\ni\perelomov. A. Perelomov, ``Generalised Coherent States and their
Applications'',
Texts and Monographs in Physics, Springer-Verlog (1986).

\ni\kikkawa. K. Kikkawa, Ann. Phys. 135, 222(1981).

\ni\trieste.  S. G. Rajeev, in "1991 Summer School in High Energy Physics and
Cosmology"  Vol. 2  ed. E. Gava, K. Narain, S. Randjbar--Daemi, E. Sezgin and
Q. Shafi, Wold Scientific (1992).

\ni\pressegal. A. Pressley and G. Segal "Loop groups", Clarendon  Press,
Oxford, (1986).

\ni\thooft. G. \' t Hooft, Nucl. Phys. B72, 461, (1974)

\ni\witten. E.Witten, Nucl. Phys. B160 (1979) 57

\ni\skyrme. T.H.R.Skyrme, Proc. Roy. Soc. A260 (1961) 127;  Nucl. Phys. 31, 556
(1962);
 J. Math. Phys. 12, 1735 (1971)

\ni\qcdskyrme. A.Balachandran, V.P.Nair, S.G.Rajeev and A.Stern, Phys. Lett.
49, 1124 (1982);  G.Adkins, C.Nappi and E.Witten, Nucl. Phys B228, 552 (1983).

\ni\rajeevetal.P. Bedaque, I. Horvath and S. G. Rajeev, Mod. Phys. Lett.A7,
3347 (1992).

\ni\chern. S.S. Chern, "Complex Manifolds without Potential Theory",
Springer--Verlag (1979).

\ni\embed. S. G. Rajeev, Phys. Rev. D44,1836, (1991).

\ni\kirillov. A. A.  Kirillov,"Elements of the Theory of Representations",
Springer--Verlag, Berlin, (1976).

\ni\arnold. V. I. Arnold, "Mathematical Methods of Classical Mechanics",
Springer--Verlag, (1978).

\ni\guest. M. A.  Guest, Proc. London Math. Soc. (3) 62, 77 (1991).

\ni\simon.  B. Simon {\it Trace Ideals and Their
Applications}
Cambridge University Press,  Cambridge,  (1979).

\ni\bars. I. Bars, Phys. Rev. Lett. 36,1521 (1976).

\ni\kp. V. Kac and  D. H. Peterson, Proc. Natl. Acad. Sci. USA 78, 3308 (1981).

\ni\lieb. E. Lieb, Phys. Rev. Lett. 46,185 (1981).

\ni\milnorch. J.  Milnor " Charecteristic Classes", Princeton University Press
(1963).

\bye

1. Definition of Infinite Dimensional Grassmannian.
Parametrization in terms of
$M(x,y)$. $\eps$ as the Hilbert transform. winding number;
cohomology.
symplectic structure. Poisson brackets. Complex structure.
Det^* bundle. Its
powers. Holomorphic sections.

2. Definition of classical hadron theory in light cone
variables. Poisson
brackets, hamiltonian and constraint. Small oscilltaions
give 't hooft
equations.

3. Quantization  by canonical methods. Role of constarint.
Why
$N_c$ is an
integer.

4. 2DQCD is equivalent to 2dhadron theory.

5. semi--classical method. meson spectrum

6. Baryon solution. small oscillations.

\bye